\documentclass[aps,prd,reprint,amsmath,amssymb,superscriptaddress,nofootinbib,eqsecnum,floatfix]{revtex4-1}

\usepackage{newtxtext}
\usepackage[varg]{newtxmath}
\usepackage[T1]{fontenc}

\IfFileExists{stylesheet.tex}{
\usepackage[stretch=20,shrink=10,final]{microtype}
\linespread{1}
\setlength{\parskip}{0pt}

\let\revtitle\maketitle
\renewcommand{\maketitle}{%
	\revtitle
	\tolerance=500
	\hyphenpenalty=1000
}

\setlength{\textfloatsep}{0.7em}
\setlength{\abovecaptionskip}{5pt}

\renewcommand{\dot}[1]{\overset{\boldsymbol{.}}{#1}\vphantom{#1}}

\DeclareFontFamily{U}{futm}{}
\DeclareFontShape{U}{futm}{m}{n}{<-> fourier-bb}{}
\DeclareMathAlphabet{\mathbb}{U}{futm}{m}{n}

\DeclareSymbolFont{cmreg}{OT1}{cmr}{m}{n}
\DeclareSymbolFont{cmmath}{OML}{cmm}{m}{i}
\DeclareSymbolFont{cmsymbols}{OMS}{cmsy}{m}{n}
\DeclareSymbolFont{cmlargesymbols}{OMX}{cmex}{m}{n}
\DeclareSymbolFontAlphabet{\mathcal}{cmsymbols}

\DeclareMathSymbol{\partial}{0}{cmmath}{64}
\DeclareMathSymbol{g}{\mathalpha}{cmmath}{103}
\DeclareMathSymbol{\eta}{0}{cmmath}{17}
\DeclareMathSymbol{\kappa}{0}{cmmath}{20}
\DeclareMathSymbol{\mu}{0}{cmmath}{22}
\DeclareMathSymbol{\nu}{0}{cmmath}{23}
\DeclareMathSymbol{\rho}{0}{cmmath}{26}
\DeclareMathSymbol{\sigma}{0}{cmmath}{27}
\DeclareMathSymbol{\ell}{0}{cmmath}{96}

\DeclareMathSymbol{\ointop}{\mathop}{cmlargesymbols}{72}
\DeclareMathSymbol{\intop}{\mathop}{cmlargesymbols}{82}
\DeclareMathDelimiter{(}{\mathopen}{cmreg}{40}{cmlargesymbols}{0}
\DeclareMathDelimiter{)}{\mathclose}{cmreg}{41}{cmlargesymbols}{1}

\DeclareMathDelimiter{(}{\mathopen}{cmreg}{40}{cmlargesymbols}{0}
\DeclareMathDelimiter{)}{\mathclose}{cmreg}{41}{cmlargesymbols}{1}
\DeclareMathDelimiter{[}{\mathopen}{cmreg}{91}{cmlargesymbols}{2}
\DeclareMathDelimiter{]}{\mathclose}{cmreg}{93}{cmlargesymbols}{3}

\usepackage{titlesec}

\newcommand{\mysectionnumbering}{\thesection.~}

\titleformat{\section}{\bfseries\center\uppercase}{\mysectionnumbering}{0em}{}
\titleformat{\subsection}{\bfseries\center}{}{0.1em}{\thesubsection.~}
\titleformat{\subsubsection}{\bfseries\itshape\center}{}{0.1em}{\thesubsubsection.~}

\titlespacing{\section}{0pt}{1.7em plus 0.9em minus 0.9em}{0.72em plus 0.3em minus 0.2em}
\titlespacing{\subsection}{0pt}{1.5em plus 0.1em minus 0.1em}{0.5em}
\titlespacing{\subsubsection}{0pt}{1.5em plus 0.1em minus 0.1em}{0.5em}

\titleformat{\paragraph}[runin]{\itshape}{}{0em}{}[.---\,]
\titlespacing{\paragraph}{\the\parindent}{0em}{0em}

\let\oldappendix\appendix

\makeatletter
\renewcommand{\appendix}{%
\@ifstar
\oneappendix
\manyappendices
}
\makeatother

\newcommand{\oneappendix}{%
\oldappendix*
\renewcommand{\mysectionnumbering}{\MakeUppercase{Appendix}:~}
\renewcommand\theequation{\Alph{section}\arabic{equation}}
}

\makeatletter
\newcommand{\manyappendices}{%
\oldappendix
\renewcommand{\mysectionnumbering}{\MakeUppercase{Appendix}~\thesection:~}
\renewcommand\theequation{\Alph{section}\arabic{equation}}
\renewcommand{\p@subsection}{\thesection}
}
\makeatother

\makeatletter
\renewcommand\@makefntext[1]{%
\noindent{\hspace{1em}}{\@makefnmark}#1}
\makeatother

\makeatletter
\renewcommand\@makefnmark{\hbox{\color{black}\@textsuperscript{\normalfont\@thefnmark}}}
\makeatother

\setlength{\skip\footins}{15pt}

\renewcommand{\footnoterule}{%
  \kern -3pt
  \hrule width 1.2cm
  \kern 4pt
}

\usepackage[dvipsnames]{xcolor}
\definecolor{revblue}{HTML}{2d3092}
\colorlet{blue}{revblue} 

\let\revcite\cite
\renewcommand\cite[1]{\mbox{\color{blue}\revcite{#1}}}

\let\reveqref\eqref
\renewcommand\eqref[1]{{\color{blue}\reveqref{#1}}}

\IfFileExists{apsrev-custom.bst}{\bibliographystyle{apsrev-custom}}{}

\newlength\mycitespacing
\settowidth\mycitespacing{\space}
\addtolength\mycitespacing{-0.5pt}
\setcitestyle{citesep={,\kern-\mycitespacing}}

}{}

\usepackage{epsfig}
\usepackage{subcaption}
\usepackage[singlelinecheck=false,justification=RaggedRight]{caption}
\usepackage{graphicx}
\usepackage{dcolumn}
\usepackage{bm}
\usepackage{makecell}
\usepackage{diagbox}
\usepackage{slashed}

\usepackage[
	hyperfootnotes=false,
	colorlinks=true,
	allcolors=blue,
	breaklinks=true]
{hyperref}

\newcommand{\be}{\begin{equation}}
\newcommand{\ee}{\end{equation}}
\newcommand{\ba}{\begin{eqnarray}}
\newcommand{\ea}{\end{eqnarray}}
\newcommand{\ecc}{{\mathfrak e}}
\newcommand{\ff}{{\mathfrak f}}
\newcommand{\NewtonG}{{\cal G}}

\begin{document}

\title{Gravitational waves from binary black holes in a self-interacting scalar dark matter cloud}

\author{Alexis Boudon}
\affiliation{Universit\'{e} Paris-Saclay, CNRS, CEA, Institut de physique th\'{e}orique, 91191, Gif-sur-Yvette, France}

\author{Philippe Brax}
\affiliation{Universit\'{e} Paris-Saclay, CNRS, CEA, Institut de physique th\'{e}orique, 91191, Gif-sur-Yvette, France}
\affiliation{CERN, Theoretical Physics Department, Geneva, Switzerland}

\author{Patrick Valageas}
\affiliation{Universit\'{e} Paris-Saclay, CNRS, CEA, Institut de physique th\'{e}orique, 91191, Gif-sur-Yvette, France}

\author{Leong Khim Wong}
\affiliation{Universit\'{e} Paris-Saclay, CNRS, CEA, Institut de physique th\'{e}orique, 91191, Gif-sur-Yvette, France}

\begin{abstract}

We investigate the imprints of accretion and dynamical friction on the gravitational-wave signals emitted by binary black holes embedded in a scalar dark matter cloud. As a key feature in this work, we focus on scalar fields with a repulsive self-interaction that balances against the self-gravity of the cloud. To a first approximation, the phase of the gravitational-wave signal receives extra correction terms at $-3$PN, $-4$PN and $-5.5$PN orders, relative to the prediction of vacuum general relativity, due to cloud gravity, accretion and dynamical friction. Future observations by LISA and DECIGO have the potential to detect these effects for a large range of scalar masses~$m_\mathrm{DM}$ and self-interaction couplings~$\lambda_4$. This would correspond to scenarios with dark matter clouds smaller than $0.1$ pc, which would be difficult to detect by other probes.

\end{abstract}

\maketitle


\section{Introduction}
\label{sec:introduction}

Perturbations to the orbits of compact objects, like black holes (BHs), can serve as a dynamical probe of their local environment. One important effect is dynamical friction, first calculated in a seminal paper by Chandrasekhar~\cite{Chandrasekhar:1943ys} for collisionless particles, and later extended to gaseous media in, e.g., Refs.~\cite{Dokuchaev1964, Ruderman1971, Rephaeli1980, Ostriker:1998fa}. These quantities were also calculated in the case of fuzzy dark matter (FDM), in the nonrelativistic and relativistic regimes \cite{Hui:2016ltb, Lancaster:2019mde, Annulli:2020lyc, Traykova:2021dua, Chowdhury:2021zik, Wang:2021udl, Vicente:2022ivh, Traykova:2023qyv}.
In this paper, we focus on the case of self-interacting dark matter, which we considered in
\cite{Brax:2019npi,Boudon:2022dxi,Boudon:2023qbu}.
In all of these cases, the compact object decelerates as it exchanges momentum with distant particles - or ``streamlines'' - that are deflected by its gravitational field. Equivalently, one can think of dynamical friction as the gravitational pull on the compact object exerted by the resulting fluid overdensity that forms in its wake. A second effect is the accretion of matter onto the~compact~object.%

Naturally, the amount of influence these effects can have on the compact object's trajectory depends on the specific nature of the environment. We are interested here in the case of dark matter clouds, within which most binary systems are expected to reside. Motivated by the lack of experimental evidence for weakly interacting massive particles (see, e.g., the reviews in Refs.~\cite{Roszkowski:2017nbc, Arcadi:2017kky}), we focus on scalar-field dark matter models with a particle mass between $10^{-20}~\text{eV}$ and $1~\text{eV}$. Within this range, very large occupation numbers are needed to form a galactic halo; hence, the scalar field behaves essentially classically and is described by a Schr\"{o}dinger wave function in the nonrelativistic regime. Static equilibrium solutions, also called ``solitons,'' form at the centers of these halos~\cite{Goodman:2000tg, Schive:2014dra, Schive:2014hza, Arbey:2001qi, Chavanis:2011zi, Chavanis:2011zm, Marsh:2015wka, Calabrese:2016hmp, Chen:2016unw, Schwabe:2016rze, Veltmaat:2016rxo, Gonzalez-Morales:2016yaf, Robles:2012uy, Bernal:2017oih, Mocz:2017wlg, Mukaida:2016hwd, Vicens:2018kdk, Bar:2018acw, Eby:2018ufi, Bar-Or:2018pxz, Marsh:2018zyw, Chavanis:2018pkx, Emami:2018rxq, Levkov:2018kau, Broadhurst:2019fsl,Hayashi:2019ynr, Bar:2019bqz, Garcia:2023abs}.
In this article, we investigate the impact on the gravitational-wave~(GW) signal emitted by a binary BH that is embedded in one of these solitons.

In the wider cosmological context, the energy density of dark matter in these scenarios is determined by the misalignment
mechanism~\cite{Preskill:1982cy, Abbott:1982af, Dine:1982ah, Arvanitaki:2019rax}, wherein the field is initially frozen but then oscillates rapidly once its mass exceeds the Hubble rate. For scalar-field potentials that are dominated by their mass term, the energy density decays as~$a(t)^{-3}$, as it does for cold dark matter~(CDM), with $a(t)$ the cosmic scale factor. One thus recovers the main predictions of the standard CDM paradigm on cosmological scales~\cite{Hu:2000ke, Johnson:2008se, Hwang:2009js, Park:2012ru, Hlozek:2014lca, Cembranos:2015oya, Urena-Lopez:2015gur, Urena-Lopez:2019kud}. Meanwhile, the details of what transpires on smaller scales depends on how strongly the scalar self-interacts. For negligible self-interactions, solitons are supported against gravitational collapse by the wavelike nature of the scalar field, which gives rise to a so-called ``quantum pressure''---this is commonly referred to as the fuzzy dark matter scenario~\cite{Hui:2016ltb}. Allowing for a repulsive, quartic interaction term introduces additional pressure effects~\cite{Goodman:2000tg, Li:2013nal, Suarez:2015fga, Suarez:2016eez, Suarez:2017mav}, however, which can even dominate over the quantum pressure in certain cases. This occurs when the soliton size is greater than the scalar's de~Broglie wavelength, and this will be the regime of interest in~this~paper.%

Solitons with radii on the order of a kiloparsec may alleviate some of the small-scale problems in galaxies encountered by the standard CDM scenario, such as the core/cusp problem, the too-big-to fail problem, or even the missing satellites problem~\cite{Hui:2001wy, deBlok:2009sp, Weinberg:2013aya, DelPopolo:2016emo}. We note, however, that other scenarios suggest that solitons could also form at higher redshifts and be of a much smaller size (see, e.g., Ref.~\cite{Brax:2020oye}). In this paper, we make no a priori assumptions about the size of the soliton, and will instead explore what information
can be extracted from GW signals for all possible values of~soliton~radii.

We consider the effects of both accretion and dynamical friction on the waveform. A BH moving inside a (much larger) soliton disturbs the distribution of dark matter both locally and further out into the bulk. Near the BH, the density of infalling dark matter grows as ${\rho \propto 1/r}$ until it reaches a nonlinear and relativistic regime close to the horizon~\cite{Brax:2019npi, Boudon:2022dxi, Boudon:2023qbu}. This inner-radius boundary condition sets the accretion rate onto the BH. At larger distances, dynamical friction arises due to the deflection of streamlines over the bulk of the scalar cloud. As for gaseous media~\cite{Dokuchaev1964, Ruderman1971, Rephaeli1980, Ostriker:1998fa}, neglecting the backreaction of the scalar field causes the dynamical friction force to vanish in the subsonic regime~\cite{Boudon:2022dxi, Boudon:2023qbu}. Both effects decrease the relative velocity between the BH and the scalar cloud. For BHs in a binary system, the consequence is a higher rate of orbital decay than if the binary were to evolve solely due to the emission of GWs. In standard post-Newtonian (PN) terminology, we find that accretion first contributes to the GW phase at the $-4$PN level for the subsonic regime and moderate supersonic Mach numbers, and at the $-5.5$PN level for high Mach numbers, while dynamical friction is a $-5.5$PN~order~effect.

The remainder of this paper is organized as follows. In Sec.~\ref{sec:eqs-of-motion}, we begin by reviewing the self-interacting model of scalar-field dark matter that we consider. In Sec.~\ref{sec:bin}, we then solve for the motion of a binary BH in the presence of a scalar cloud. The perturbations to the phase of the emitted GWs arising from accretion and dynamical friction are derived in Sec.~\ref{sec:phase}.
We describe our Fisher-matrix analysis in Sec.~\ref{sec:fish} and finally, in Sec.~\ref{sec:detection} we forecast
the prospects of detecting such a dark matter environment in current and future GW experiments.
We conclude in Sec.~\ref{sec:conclusion}.

\section{Equations of motion}
\label{sec:eqs-of-motion}

\subsection{Scalar field dark matter}

In this paper, we study the signatures imprinted on the gravitational waveform of a binary system of BHs by dark matter environments associated with a self-interacting scalar field. The dynamics of the scalar are governed by the action
\be
S_\phi = \int \frac{d^4 x}{\hbar c^2} \sqrt{-g} \left( - \frac{1}{2} g^{\mu\nu} \partial_\mu\phi
\partial_\nu\phi - V(\phi) \right),
\label{eq:Action}
\ee
where we take the scalar-field potential to be
\be
V(\phi) = \frac{m_{\rm DM}^2 c^2}{2 \hbar^2} \phi^2 +  \frac{\lambda_4}{4 \hbar^2 c^2} \phi^4,
\label{eq:V_I-def}
\ee
with coupling constant ${\lambda_4 > 0}$. This gives rise to a repulsive self-interaction
between dark matter particles in the nonrelativistic limit, wherein
the global behavior of dark matter is akin to that of a compressible fluid. The effective outward pressure of this repulsive interaction can counterbalance the attractive force of gravity, and therefore leads to the formation of stable, equilibrium dark matter configurations on small scales, called~solitons.%

A detailed cosmological analysis of this dark matter model is presented in Ref.~\cite{Brax:2019fzb}.
We here briefly review the main points.
On cosmological scales, the oscillations of the scalar field due to the
quadratic mass term in $V(\phi)$ are dominant since at least the time of matter-radiation equality. This ensures that
the scalar field behaves as dark matter with a background density~$\bar\rho$ that decays with the scale factor~$a(t)$ as
${\bar\rho \propto a(t)^{-3}}$.
However, the pressure associated with the self-interaction term prevents the growth of
density perturbations below the Jeans scale
\be
r_a = \frac{c}{\sqrt{4\pi\NewtonG \rho_a}} , \;\;\; \rho_a = \frac{4 m_{\rm DM}^4 c^3}{3\lambda_4 \hbar^3} .
\label{eq:ra-def}
\ee
The characteristic scale $r_a$ actually sets both the cosmological Jeans length, which
leads to a small-scale cutoff for cosmological structure formation, and the radius of the soliton~\cite{Goodman:2000tg, Chavanis:2011uv}.

In the nonrelativistic regime, the nonlinear Klein-Gordon equation derived from the action
in Eq.~\eqref{eq:Action} reduces to the nonlinear Schr\"odinger-Poisson system. In simple configurations
(wherein the density does not vanish), a Madelung transformation~\cite{Madelung1926}
can be used to map this onto a hydrodynamical system, in which case the solitons correspond to hydrostatic equilibria.
The quartic self-interaction in Eq.~\eqref{eq:V_I-def} gives rise to an effective pressure
$P \propto \rho^2$, not unlike a polytropic gas with index ${\gamma=2}$. The soliton density profile
then takes~the~form
\be
\rho_{\rm sol}(r) = \rho_0  \frac{\sin(\pi r/R_{\rm sol})}{\pi r/R_{\rm sol}},
\quad R_{\rm sol} = \pi r_a,
\label{eq:rho-sol-TF}
\ee
in the Thomas-Fermi limit of negligible quantum pressure.
Observe that such solitons are described by just three parameters: the fundamental constants $m_{\rm DM}$ and~$\lambda_4$, and the average bulk density~$\rho_0$. The value of this last quantity---or, equivalently, the value of the soliton mass $M_{\rm sol}=(4/\pi)\rho_0 R_{\rm sol}^3$---depends on the formation history of the dark matter halo.

If the characteristic scale~$r_a$ in Eq.~\eqref{eq:ra-def} is on the order of a kiloparsec or more, then these solitons form at the centers of galaxies, as in the FDM case~\cite{Schive2014a}, while the outer regions of the dark matter halo follow an NFW density profile~\cite{Navarro:1995iw}. A numerical study of such soliton-halo systems for the potential in Eq.~\eqref{eq:V_I-def} is presented in Ref.~\cite{Garcia:2023abs}. On scales greater than $R_{\rm sol}$ and the de~Broglie wavelength ${\lambda_{\rm dB} \equiv 2\pi \hbar /(m_{\rm DM} v)}$, both the self-interaction and quantum pressure are negligible, and so scalar-field dark matter behaves as collisionless cold dark matter would.
Moreover, even though $r_a$ is fixed, increasingly large and massive halos can form hierarchically in this model, as in the standard CDM paradigm~\cite{1980lssu.book.Peebles}.

At the other end of the spectrum, if $r_a$ is much smaller than the typical size of galaxies, then solitons may have
formed at early times before the formation of galaxies. In a manner similar to the formation of primordial BHs, this could lead to macroscopic dark matter objects with radii ranging from that of an asteroid to giant molecular clouds~\cite{Brax:2020oye}.
Indeed, if the hierarchy of scales is sufficiently large, then many small solitons may be present within galactic
halos. In this scenario, stellar-mass binary BH systems could happen to be embedded within such
solitons. We shall investigate the impact of both types of solitons---galactic sized or smaller---on the motion of binary~BHs.

Several assumptions have been made to render the calculations in this paper feasible. First, note that the sound speed of the dark matter fluid is given by~\cite{Brax:2019npi, Boudon:2022dxi}
\be
c_s^2 = \frac{\rho_0}{\rho_a} c^2,
\label{eq:cs2-def}
\ee
as would be expected for a polytropic gas with index ${\gamma=2}$. We restrict ourselves to the nonrelativistic regime wherein ${c_s \ll c}$, and thus~${\rho_0 \ll \rho_a}$. We further limit our attention to the large-scalar-mass limit,
\be
m_\text{DM} > \frac{\hbar}{r_s c} = 7 \times 10^{-11} \left( \frac{m_{\rm BH}}{1 \, M_\odot} \right)^{-1}~\text{eV},
\label{eq:large-m-rs}
\ee
where ${r_s \equiv 2\NewtonG m_\text{BH} / c^2}$ is the Schwarzschild radius of the larger of the two BHs embedded in the soliton. Taking this limit amounts to assuming that the scalar's de~Broglie and Compton wavelengths are smaller than the BH's horizon, and much smaller than the size of the soliton. The analytic formulas for the accretion rate and dynamical friction force that we use below were derived in Refs.~\cite{Brax:2019npi, Boudon:2022dxi, Boudon:2023qbu} and are valid only when this holds. Conveniently, a by-product of this assumption is that the only dark matter parameters affecting the binary's motion are the two characteristic densities, $\rho_a$~and~$\rho_0$.%

As a BH moves inside such dark matter solitons, it slows down because of
the accretion of dark matter and the dynamical friction with the dark matter environment.
In addition, it feels the gravitational potential of the dark matter cloud.
We describe these effects in the next three sections.

\subsection{Accretion drag force}

For the particular model in Eqs.~\eqref{eq:Action} and~\eqref{eq:V_I-def}, it was shown in Ref.~\cite{Boudon:2023qbu} that the accretion rate of scalar dark matter onto a BH follows
two regimes,
\be
v_{\rm BH} < v_{\rm acc} : \;
\dot m_{\rm BH} = \dot m_{\rm max} , \;\;\;
v_{\rm BH}> v_{\rm acc} : \; \dot m_{\rm BH} = \dot m_{\rm BHL} ,
\label{eq:accretion-rate}
\ee
with
\ba
&& v_{\rm acc} = \frac{c_s^{2/3} c^{1/3}}{(3 F_\star)^{1/3}} , \;\;\;
\dot m_{\rm max} = 3\pi F_\star \rho_a r_s^2 c =
\frac{12 \pi F_\star \rho_0 \NewtonG ^2 m_{\rm BH}^2}{c_s^2 c} , \nonumber \\
&& \dot m_{\rm BHL} = \frac{4 \pi \rho_0 \NewtonG ^2 m_{\rm BH}^2}{v_{\rm BH}^3} ,
\label{eq:v-acc}
\ea
where an overdot denotes differentiation with respect to time and ${F_\star \simeq 0.66}$
is obtained from a numerical computation of the critical flux \citep{Brax:2019npi},
which is associated with the unique radial transonic solution that matches the supersonic
infall at the Schwarzschild radius to the static equilibrium soliton at large distances.
This critical behavior is similar to that found for hydrodynamical flows in the classic studies
of Refs.~\cite{Bondi:1952ni, Michel:1972}, and is closely related to the case of a polytropic
gas with index $\gamma=2$~\cite{Brax:2019npi,Boudon:2022dxi}. However, close to the BH, the
dynamics deviates from that of a polytropic gas as one enters the relativistic regime.
Near the Schwarzschild radius, the scalar field must be described by the nonlinear
Klein-Gordon equation instead of hydrodynamics~\cite{Brax:2019npi}. This implies that the
critical flux and the accretion rate $\dot m_{\rm max}$ differ from the usual
Bondi result ${\dot m_{\rm Bondi} \sim \rho_0 \NewtonG ^2 m_{\rm BH}^2/c_s^3}$.
This is manifest in the dependence of $\dot m_{\rm max}$ on the speed of light~$c$,
which is absent from the usual Bondi result.%

The high-velocity regime corresponds to the standard accretion-column picture
\citep{1939PCPS...35..405H,Bondi-Hoyle-1944} and we recover the Bondi-Hoyle-Lyttleton
accretion rate $\dot m_{\rm BHL}$. There, most of the accretion
comes from the narrow wake behind the BH, delimited by a conical shock within the Mach
angle $\sin\theta_c = 1/{\cal M}\ll 1$, where ${\cal M}=v_{\rm BH}/c_s$ is the BH Mach number.

In the low-velocity regime the Bondi-Hoyle-Lyttleton accretion rate is greater than the
maximum accretion rate $\dot m_{\rm max}$ that is allowed by the effective pressure
associated with the self-interactions (close to the BH horizon the velocity cannot be greater
than $c$ and the density greater than $\rho_a$). Then, the accretion column is no longer
a narrow cone behind the BH and it encloses the BH from all sides. There is a bow shock
upstream of the BH, with a subsonic region that contains the BH and diverts most of the
dark matter flux. Close to the horizon the flow is approximately radial and we recover
the accretion rate $\dot m_{\rm max}$. See \cite{Boudon:2023qbu} for details.

Now consider a BH moving with velocity ${\bf v}_{\rm BH}$ through this scalar cloud.
In the nonrelativistic limit $v_\text{BH} \equiv |{\bf v}_{\rm BH}| \ll c$ and in the
reference frame of the cloud, the accretion of zero-momentum dark matter does not change
the BH momentum but slows down its velocity as
\be
m_{\rm BH} \dot{\bf v}_{\rm BH} |_{\rm acc}= - \dot m_{\rm BH} {\bf v}_{\rm BH} .
\label{eq:dvdt-acc}
\ee

\subsection{Dynamical friction}

Dynamical friction also acts to reduce the BH's velocity. As in the hydrodynamical
case~\cite{Dokuchaev1964, Rephaeli1980, Ostriker:1998fa}, the dynamical friction force
(in the steady-state limit) vanishes for subsonic speeds
${v_{\rm BH} < c_s}$~\cite{Boudon:2022dxi} but is nonzero at supersonic speeds.
The additional force on the BH in the latter regime reads~\cite{Boudon:2023qbu}
\be
m_{\rm BH} \dot{\bf v}_{\rm BH} |_{\rm df} = - \frac{8\pi \NewtonG ^2 m_{\rm BH}^2 \rho_0}
{3 v_{\rm BH}^3} \ln \left( \frac{r_{\rm IR}}{r_{\rm UV}} \right) {\bf v}_{\rm BH} ,
\label{eq:DF}
\ee
where $r_{\rm IR}$ is the usual large-radius cutoff while the small-radius cutoff of
the logarithmic Coulomb factor is given by
\be
r_{\rm UV} = 3 \sqrt{\frac{2}{e}} r_{\rm sg} {\cal M}^{-3/2}
= 6 \sqrt{\frac{2}{e}} \frac{\NewtonG  m_{\rm BH}}{c_s^2}
\left( \frac{c_s}{v_{\rm BH}} \right)^{3/2} .
\label{eq:r-UV}
\ee
Here $e$ is Euler's number (not to be confused with the orbital eccentricity~$\ecc$
in Sec.~\ref{sec:bin}), ${\cal M} = v_{\rm BH}/c_s$ is the Mach number, and
$r_{\rm sg}=r_s c^2/c_s^2$.
Equation~\eqref{eq:DF} takes the same form as the collisionless result by
Chandrasekhar~\cite{Chandrasekhar:1943ys} but with a multiplicative factor $2/3$.
It is not so surprising to obtain a result that differs from Chandrasekhar's formula,
even for distant streamlines. Indeed, the background made of the soliton is governed
by the balance between gravity and self-interactions, so that the self-interactions are never
negligible throughout the dark matter soliton. We can also note that in the subsonic regime,
the dynamical friction is zero, which shows the global impact of the self-interactions
(which generate the sound speed) throughout the medium, in the steady state.
Finally, in the collisionless case, distant trajectories that are deflected by small angles
would nevertheless cross each other along the symmetry axis at large distance behind the BH,
which is not possible for a fluid with non-zero self-interactions. Therefore, even distant
streamlines must depart from distant collisionless trajectories.
These various arguments explain why we could expect a different result from
Chandrasekhar's formula even for distant streamlines (as long as they remain within
the dark matter soliton).

In addition, the ultra-violet cutoff $r_{\rm UV}$ is here fully determined by the physics
of the scalar field and its effective pressure, instead of the minimum impact parameter
${b_{\rm min} \sim \NewtonG  m_{\rm BH}/v_{\rm BH}^2}$.
As we have $r_{\rm UV} \sim b_{\rm min}  \sqrt{v_{\rm BH}/c_s} > b_{\rm min}$,
we can see that the dynamical friction (\ref{eq:DF}) is smaller than the collisionless
result, with a damping factor below $2/3$.

The radius $r_{\rm sg}=r_s c^2/c_s^2$ in Eq.(\ref{eq:r-UV}) is the radius where in the
spherical accretion case the dark matter density profile makes the transition from
the constant large-distance value $\rho_0$ to the $1/r$ growth close to the BH.
As could be expected, $r_{\rm UV}$ decreases in units
of $r_{\rm sg}$ for smaller $c_s$ (equivalently, smaller $\lambda_4$). This falls off as
${\cal M}^{-3/2}=c_s^{3/2} v_{\rm BH}^{-3/2}$. Not surprisingly, we have $r_{\rm UV} \sim r_{\rm sg}$
for Mach numbers of the order of unity.
On the other hand, at fixed $\rho_0$, the radius $r_{\rm sg}$ grows for smaller $\lambda$
and smaller $c_s$. This is because the smaller self-interaction requires a higher density
for the pressure to be able to regulate the infall onto the BH. Therefore, in the Bondi-like
steady-state a smaller $\lambda$ leads to a higher density in the inner region and
to a transition to the constant-density plateau that is pushed to larger distance.
The growth of $r_{\rm sg}$ happens to be steeper than the factor ${\cal M}^{-3/2}$
and leads to an increase of $r_{\rm UV}$.
This expression is actually fully determined by the large-distance perturbative expansion
presented in Sec.III of Ref.~\cite{Boudon:2023qbu}.

For a steady straight-line trajectory, we may take for the infra-red cutoff the
size of the dark matter soliton, which depends explicitly on $m_{\rm DM}$ and
$\lambda_4$ via Eq.~\eqref{eq:ra-def}.
However, for bodies moving on circular orbits of radius $r_{\rm orb}$, numerical
simulations and analytical studies find that for gaseous media a good match is obtained
by using $r_{\rm IR} = 2 r_{\rm orb}$ \citep{Kim:2007zb,Desjacques:2022}.
This can be understood as follows. Estimating the dynamical friction from the exchange
of momentum with distant encounters or streamlines of impact parameter $b$, as in the
classical study \citep{Chandrasekhar:1943ys}, the duration an encounter is
$\Delta t \sim b/v_{\rm BH}$. Requiring this time to be smaller than the orbital period
$P_{\rm orb} \sim r_{\rm orb}/v_{\rm BH}$, so that the BH does not turn around during the encounter,
gives $b \lesssim r_{\rm orb}$.
If we estimate the dynamical friction from the gravitational attraction by the BH wake,
at large distance in the BH rest-frame matter flows away at the radial velocity
$v_{\rm BH}$. Therefore, the wake is aligned behind the BH up to the distance
$d \sim v_{\rm BH} P_{\rm orb}/2$, which gives again the large-radius cutoff $d \lesssim r_{\rm orb}$.
Therefore, we take
\be
r_{\rm IR} = 2 r_{\rm orb} ,
\label{eq:r-IR}
\ee
with the same normalization as found for gaseous media \citep{Kim:2007zb}.
As shown in Sec.~\ref{sec:detection} below, it turns out that the impact of the dark matter environment
on the gravitational waves signal is dominated by the accretion rather than the dynamical friction.
Therefore, our results are not very sensitive to the precise value of the infra-red cutoff (\ref{eq:r-IR}).

\subsection{Dark matter halo}

Approximating the bulk of the soliton as a spherical halo of density $\rho_0$ and radius
$R_{\rm sol}$, centered at position ${\bf x}_0$, the halo gravitational potential reads
\be
| {\bf x} - {\bf x}_0 | < R_{\rm sol} : \;\;\;
\Phi_{\rm halo}({\bf x}) = \frac{2\pi}{3} \NewtonG  \rho_0 |{\bf x}-{\bf x}_0|^2 .
\ee
This gives the gravitational acceleration
\be
m_{\rm BH} \dot{\bf v}_{\rm BH} |_{\rm halo} = - \frac{4\pi}{3} \NewtonG m_{\rm BH} \rho_0
({\bf x}-{\bf x}_0) .
\label{eq:halo-force}
\ee

\section{Binary motion}
\label{sec:bin}

We focus on a binary system of two BHs and study their dynamics in their inspiralling phase
in the Newtonian regime. Then, the Keplerian orbital motion is perturbed by the dark matter
accretion, the dynamical friction and the halo gravity, and by the emission of GWs.
This leads to a shrinking of the BH separation, until their merging.
In the large-distance inspiralling phase, we obtain the perturbations of the Keplerian
motion at first order. This allows us to consider separately the impact of the scalar
cloud and of the GWs.

\subsection{Keplerian motion}

To compute the perturbation of the orbits at first order, we use the standard method of
osculating orbital elements \cite{poisson_will_2014}, where we derive the drift of the
orbital elements that determine the shape of the orbits.
To define our notations, we first recall the properties of the Keplerian orbits.
At zeroth order, the binary system of the two BHs of masses $\{m_1,m_2\}$, positions
$\{ {\bf x}_1, {\bf x}_2 \}$ and velocities $\{ {\bf v}_1, {\bf v}_2 \}$, is reduced
to a one-body problem by introducing the relative distance ${\bf r}$,
\be
{\bf r} = {\bf x}_1 - {\bf x}_2 , \;\; {\bf v} = {\bf v}_1 - {\bf v}_2 ,
\ee
the total and reduced masses
\be
m= m_1 + m_2 , \;\; \mu = m_1 m_2 /m .
\ee
This gives the equation of motion
\be
\ddot {\bf r} = - \frac{\NewtonG  m}{r^3} {\bf r}
\ee
for the relative separation, whereas the center of mass remains at rest if its initial velocity vanishes.
Then, we also have
\be
{\bf x}_1 = \frac{m_2}{m} {\bf r} , \;\;\; {\bf x}_2 = - \frac{m_1}{m} {\bf r} , \;\;\;
{\bf v}_1 = \frac{m_2}{m} {\bf v} , \;\; {\bf v}_2 = - \frac{m_1}{m} {\bf v} ,
\label{eq:x1-x2-r}
\ee
choosing for the origin of the coordinates the barycenter of the binary system.
The solution for bound orbits is the ellipse given by
\be
r = \frac{p}{1+{\ecc} \cos(\phi-\omega)} , \;\;\; p = (1-\ecc^2) a ,
\ee
where $p$ is the orbit semi-latus rectum, $a$ the semi-major axis, $\ecc$ the eccentricity
and $\omega$ the longitude of the pericenter.
The orbit takes place in the plane $({\bf e}_x,{\bf e}_y)$ orthogonal to the axis ${\bf e}_z$.
In spherical coordinates, the polar angle $\theta=\pi/2$ is constant while the azimuthal angle $\phi$ runs.
The total angular momentum ${\bf L}$ is constant,
\be
{\bf L} = m_1 {\bf x}_1 \times {\bf v}_1 + m_2 {\bf x}_2 \times {\bf v}_2 = \mu {\bf h} ,
\ee
with
\be
{\bf h} = {\bf r} \times {\bf v} = h \, {\bf e}_z , \;\;\; h = r^2 \dot\phi , \;\; p = \frac{h^2}{\NewtonG  m} .
\ee
The constancy of $\omega$ is related to the conservation of the Runge-Lenz vector,
\be
{\bf A} = \frac{{\bf v} \times {\bf h}}{\NewtonG m} - {\bf e}_r = \ecc
(\cos\omega \, {\bf e}_x + \sin\omega \,{\bf e}_y ) .
\ee
In the following, we will also use the true anomaly defined by
\be
\varphi = \phi - \omega ,
\ee
which measures the azimuthal angle from the direction of pericenter and grows with time as
\be
\dot\varphi = \sqrt{ \frac{\NewtonG  m}{p^3} } (1 + \ecc \cos\varphi)^2 \; .
\label{eq:dphi-dt}
\ee
The period $P_{\rm orb}$ and the frequency $f_{\rm orb}$ of the orbital motion read
\be
P_{\rm orb} = 2\pi \sqrt{ \frac{a^3}{\NewtonG m} } , \;\;\;
f_{\rm orb} = \frac{1}{2\pi} \sqrt{ \frac{\NewtonG m}{a^3} } ,
\label{eq:P-period}
\ee
which is known as Kepler's third law.

\subsection{Drag force from the dark matter}

As seen in Sec.~\ref{sec:eqs-of-motion}, the equations of motion of the two BHs read
\ba
&& m_1 \ddot {\bf x}_1 = \NewtonG  m_1 m_2 \frac{ {\bf x}_2 - {\bf x}_1 }
{ | {\bf x}_2 - {\bf x}_1 |^3} - \dot{m}_1 \dot {\bf x}_1 - f_1 \dot {\bf x}_1
- g_1 ({\bf x}_1-{\bf x}_0) , \nonumber \\
&&
m_2 \ddot {\bf x}_2 =  \NewtonG  m_1 m_2 \frac{ {\bf x}_1 - {\bf x}_2 }
{ | {\bf x}_1 - {\bf x}_2 |^3} - \dot{m}_2 \dot {\bf x}_2 - f_2 \dot {\bf x}_2
- g_2 ({\bf x}_2-{\bf x}_0) , \nonumber \\
&&
\ea
where we take into account the Newtonian gravity of the binary, the accretion of dark matter,
the dynamical friction and the halo gravity, with
\be
f_i(t) = \Theta_{{\rm df}.i} \frac{8\pi \NewtonG ^2 m_i^2 \rho_0}
{3 v_i^3} \ln \left( \frac{r_{{\rm IR},i}}{r_{{\rm UV},i}} \right) , \;\;\;
g_i = \frac{4\pi}{3} \NewtonG  m_i \rho_0 .
\label{eq:fi-gi}
\ee
Here $\Theta_{{\rm df},i}$ is a Heaviside factor associated with the two conditions
$v_i > c_s$ and $r_{{\rm IR},i}>r_{{\rm UV},i}$.
This is only an approximation, however,
as a perturbative treatment to higher orders, which takes the scalar field's backreaction
onto the BH into account, should smooth out the transition at $c_s$ and give a small but
nonzero force in the subsonic regime~\cite{Berezhiani:2019pzd}. Nevertheless, we expect our
use of a sharp transition to provide a conservative estimate for the impact of the dynamical
friction on the motion~of~a~BH.

This gives for the separation ${\bf r}$ the equation of motion
\be
\ddot {\bf r} = - \frac{\NewtonG  m}{r^3} {\bf r} - \left( \frac{\dot\mu}{\mu}
+ \frac{m_2 f_1}{m_1 m} + \frac{m_1 f_2}{m_2 m} \right) \dot {\bf r}
- \frac{4\pi \NewtonG \rho_0}{3} {\bf r} .
\label{eq:ddot-r-all}
\ee
Here we used Eq.(\ref{eq:x1-x2-r}) to express ${\bf x}_i$ in terms of ${\bf r}$ in the last
two terms, as we work at first order in the perturbations $\dot m_i$, $f_i$ and $g_i$.
Thus, we obtain an equation of motion of the form
\be
\ddot {\bf r} = - \frac{\NewtonG  m(t)}{r^3} {\bf r} - F(t) \dot {\bf r} - G {\bf r} .
\label{eq:ddot-r-F}
\ee
Here and in the following, we assumed that at zeroth-order the center of mass of the binary
is at rest in the scalar cloud, or more generally that its velocity is small as compared
with the binary orbital velocity ${\bf v}$.

For circular orbits with $v=\sqrt{\NewtonG  m/a}$, we obtain
\be
\frac{r_{{\rm IR},i}}{r_{{\rm UV},i}} = \sqrt{\frac{e c_s m^2 \mu^5}{18 v m_i^7}} , \;\;
\frac{v_i}{c_s} = \frac{\mu v}{m_i c_s}
\ee
and the Heaviside factor in Eq.(\ref{eq:fi-gi}) reads
\be
\Theta_{{\rm df},i} = \Theta\left( \frac{m_i}{\mu} < \frac{v}{c_s} < \frac{e m^2 \mu^5}{18 m_i^7} \right) ,
\ee
which is unity when the conditions are satisfied and zero otherwise.
We can see that the conditions $r_{{\rm IR},i}>r_{{\rm UV},i}$ and
$v_i>c_s$ can only be simultaneously satisfied by the smallest BH of the binary,
when the symmetric mass ratio $\nu$ defined by
\be
\nu= \mu/m = m_1 m_2/m^2
\label{eq:nu-def}
\ee
is below
\be
\nu \lesssim 0.16 .
\ee

Following the method of the osculating orbital elements \cite{poisson_will_2014}, we obtain the
impact of the accretion and of the dynamical friction by computing the perturbations to the orbital
elements. It is clear from Eq.(\ref{eq:ddot-r-F}) that the orbital plane remains constant.
In particular, the specific angular momentum ${\bf h}$ remains parallel to ${\bf e}_z$ and evolves
as
\be
\dot {\bf h} = - F(t) {\bf h} ,
\ee
whereas the Runge-Lenz vector evolves as
\be
\dot {\bf A} =  - \left( \frac{\dot m}{m} + 2 F(t) \right) ( {\bf A} + {\bf e}_r ) + \frac{G h r}{\NewtonG m} {\bf e}_\phi.
\label{eq:dAdt-general}
\ee
This gives next the evolution of the eccentricity and of the semi-major axis,
\ba
&& \hspace{-0.5cm} \dot \ecc = - \left( \frac{\dot m}{m} + 2 F(t) \right) ( \ecc+\cos\varphi)
- \frac{G h a (1-\ecc^2) \sin\varphi}{\NewtonG m (1+\ecc\cos\varphi)} , \nonumber \\
&& \hspace{-0.5cm} \dot a = - \left( \frac{\dot m}{m} + 2 F(t) \right) \frac{a ( 1 + \ecc^2 + 2 \ecc \cos\varphi )}
{1-\ecc^2} - \frac{2 G h \ecc a^2 \sin\varphi}{\NewtonG m (1+\ecc \cos\varphi)} . \nonumber \\
&&
\label{eq:dot-e-dot-a}
\ea
Using Eq.(\ref{eq:dphi-dt}), the derivatives with respect to the true anomaly $\varphi$ read
at first order
\ba
\frac{d\ecc}{d\varphi} & = & - \sqrt{ \frac{p^3}{\NewtonG  m} } \biggl \lbrace \left( \frac{\dot m}{m} + 2 F(t) \right)
\frac{\ecc+\cos\varphi}{(1+\ecc \cos\varphi)^2} \nonumber \\
&& + \frac{G h a (1-\ecc^2)}{\NewtonG m} \frac{\sin\varphi}{(1+\ecc\cos\varphi)^3} \biggl \rbrace
\label{eq:ecc-F}
\ea
and
\ba
\frac{da}{d\varphi} & = & - \sqrt{ \frac{p^3}{\NewtonG  m} } \biggl \lbrace \left( \frac{\dot m}{m} + 2 F(t) \right)
\frac{a}{1-\ecc^2} \frac{1 + \ecc^2 + 2 \ecc \cos\varphi}{(1+\ecc \cos\varphi)^2} \nonumber \\
&& + \frac{2 G h \ecc a^2}{\NewtonG m} \frac{\sin\varphi}{(1+\ecc\cos\varphi)^3} \biggl \rbrace .
\label{eq:a-F}
\ea
The perturbations generated by the dark matter lead to oscillations and secular changes of the orbital
elements. The cumulative drift associated with the secular effects is obtained by averaging over one
orbital period, as
\be
\langle \dot a \rangle = \frac{1}{P} \int_0^P dt \, \dot a = \frac{1}{P} \int_0^{2\pi} d\varphi \, \frac{da}{d\varphi} .
\label{eq:average-P}
\ee

\subsection{Effect of the accretion}
\label{sec:impact-acc}

We first consider the impact of the accretion of dark matter on the orbital motion.
This corresponds to both the term $\dot m/m$ and the contribution
$F_{\rm acc} = \dot \mu/\mu$ to $F(t)$.
We focus on the regime where these accretion rates vary slowly as compared with the orbital motion
and we take them constant over one period.
As seen in (\ref{eq:accretion-rate}), we have two regimes for the accretion rates, which are constant at low
velocity and decays as $v_i^{-3}$ at high velocity.
Thus, we can write
\be
\frac{\dot m}{m} + 2 \frac{\dot\mu}{\mu} = A_{\rm acc} + \frac{B_{\rm acc}}{v^3} ,
\ee
with
\ba
&& \hspace{-0.5cm} A_{\rm acc} =  \frac{12\pi F_\star \NewtonG^2 \rho_0 \mu}{c_s^2 c} \sum_{i=1}^2
\Theta(v_i < v_{\rm acc}) \left( 2+ \frac{m_i^2}{m \mu} \right) , \nonumber \\
&& \hspace{-0.5cm} B_{\rm acc} = 4\pi \NewtonG ^2\rho_0 \mu \sum_{i=1}^2
\Theta(v_i > v_{\rm acc}) \frac{m_i^3}{\mu^3} \left( 2+ \frac{m_i^2}{m \mu} \right) .
\label{eq:A-acc-B-acc-def}
\ea
Then, at lowest order over the eccentricity $\ecc$ we obtain from Eqs.(\ref{eq:ecc-F})-(\ref{eq:a-F})
\ba
&& \langle \dot\ecc \rangle_{\rm acc} = \frac{3 \ecc}{2} \left( \frac{a}{\NewtonG m} \right)^{3/2} B_{\rm acc}
, \;\;\; \nonumber \\
&& \langle \dot a \rangle_{\rm acc} = - a A_{\rm acc}  - a \left( \frac{a}{\NewtonG m} \right)^{3/2} B_{\rm acc} .
\label{eq:ecc-a-accretion}
\ea
The eccentricity remains constant in the low-velocity regime and increases in the high-velocity regime,
if $\ecc > 0$. The size of the orbit always decreases.
The result (\ref{eq:ecc-a-accretion}) for the semi-major axis can be recovered
at once for circular orbits from the constancy of the total angular momentum $L= \mu \sqrt{\NewtonG  m p}$,
with $a=p$ and $v=\sqrt{\NewtonG m/a}$ for $\ecc=0$.

\subsection{Effect of the dynamical friction}
\label{sec:impact-DF}

The dynamical friction corresponds to the contribution
\be
F_{\rm df} = \frac{m_2 f_1}{m_1 m} + \frac{m_1 f_2}{m_2 m} ,
\ee
and we can write
\be
2 F_{\rm df}(t) = \frac{B_{\rm df}}{v^3} + \frac{C_{\rm df}}{v^3} \ln\left( \frac{v}{c_s} \right) ,
\label{eq:F-DF}
\ee
with
\ba
B_{\rm df} & = & \frac{8\pi \NewtonG ^2\rho_0 \mu}{3} \sum_{i=1}^2 \Theta_{{\rm df},i}
\frac{m_i^3}{\mu^3} \ln\left(\frac{e m^2\mu^5}{18 m_i^7}\right) , \nonumber \\
C_{\rm df} & = & - \frac{8\pi \NewtonG^2\rho_0\mu}{3} \sum_{i=1}^2 \Theta_{{\rm df},i} \frac{m_i^3}{\mu^3} .
\ea
At lowest order over the eccentricity $\ecc$ we obtain
\ba
&& \langle \dot\ecc\rangle_{\rm df} = \frac{3 \ecc}{2} \left( \frac{a}{\NewtonG  m}\right)^{3/2} \left[
B_{\rm df} + C_{\rm df} \ln\left( \sqrt{\frac{\NewtonG m}{a}} \frac{1}{c_s}\right) - \frac{C_{\rm df}}{3} \right] ,
\nonumber \\
&& \langle \dot a\rangle_{\rm df} = - a  \left( \frac{a}{\NewtonG  m}\right)^{3/2} \left[
B_{\rm df} + C_{\rm df} \ln\left( \sqrt{\frac{\NewtonG m}{a}} \frac{1}{c_s} \right) \right] .
\label{eq:a-dot-df}
\ea
Thus, the dynamical friction increases the eccentricity, if $\ecc > 0$, and reduces the size of the orbit.

\subsection{GWs emission for the Keplerian dynamics}

As is well known, the emission of GWs makes the orbits become more circular
and tighter, until the BHs merge.
At lowest order in a post-Newtonian expansion and using the quadrupole formula,
the drifts of the eccentricity and of the semi-major axis are given by the standard
results \cite{poisson_will_2014}
\be
\langle \dot{\ecc} \rangle_{\rm gw} = - \frac{304 \nu c}{15 a} \ecc
\left(\frac{\NewtonG  m}{c^2 a}\right)^3 (1-\ecc^2)^{-5/2} \left(1+\frac{121}{304}\ecc^2\right)
\label{eq:ecc-GW}
\ee
and
\be
\langle \dot a \rangle_{\rm gw} = - \frac{64 \nu c}{5} \left(\frac{\NewtonG  m}{c^2 a}\right)^3
\frac{1+\frac{73}{24}\ecc^2+\frac{37}{96}\ecc^4}{(1-\ecc^2)^{7/2}} .
\label{eq:a-GW}
\ee
As pointed out in Ref.~\cite{Cardoso_2021}, at large distances the increase of eccentricity by accretion
and dynamical friction in high-density environments can lead to significant eccentricity for some
binaries as they enter the LISA observational band. This effect is somewhat lessened in our case
as the dynamical friction vanishes in the subsonic regime. In this paper, we focus on the later inspiral
stage where the impact of the dark matter on the binary is smaller than that of the emission of GWs
and we restrict ourselves to circular orbits with $\ecc = 0$.
The analysis of binaries that would have acquired a high eccentricity at earlier stages, as studied
in \cite{Cardoso_2021}, is left for a future work.

\subsection{Effect of the halo gravity}
\label{sec:impact-halo}

As can be checked at once in Eqs.(\ref{eq:ecc-F})-(\ref{eq:a-F}), the $G$-term associated with the halo
gravity does not modify the eccentricity and the size of the orbit over one period,
$\langle \dot\ecc \rangle_{\rm halo} = 0$ and $\langle \dot a \rangle_{\rm halo} = 0$.
Indeed, within the approximation (\ref{eq:halo-force}) of a time-independent halo
gravitational potential, this is a conservative force.
However, this modification of the Keplerian potential induces a change of the orbital frequency
and of the emission of gravitational waves.
Focusing on the binary and halo gravity only, the equation of motion (\ref{eq:ddot-r-all})
corresponds to the energy
\be
E = \frac{1}{2} \mu v^2 - \frac{\NewtonG \mu m}{r} + \frac{2\pi \NewtonG \rho_0 \mu r^2}{3} .
\ee
Writing the Euler-Lagrange equations of motion, we obtain for circular orbits of radius $a$
the velocity
\be
v_\phi = \sqrt{\frac{\NewtonG m}{a}} \left( 1 + \frac{2\pi\rho_0 a^3}{3 m} \right) .
\ee
Here and in the following, we work at linear order in $\rho_0$.
Thus, relative corrections to the Keplerian results are set by the ratio between the
dark matter mass inside the orbital radius and the binary total mass,
The orbital frequency and the energy read as
\be
f_{\rm orb} = \frac{1}{2\pi} \sqrt{\frac{\NewtonG m}{a^3}}
\left( 1 + \frac{2\pi\rho_0 a^3}{3 m} \right)
\label{eq:f-orb-DM}
\ee
and
\be
E = - \frac{\NewtonG m \mu}{2 a} + \frac{4\pi \NewtonG \rho_0 \mu a^2}{3} .
\label{eq:E-circular-halo}
\ee
As expected, the higher mass in the system, and hence the larger gravity, increases the
orbital frequency. Using the quadrupole formula \citep{poisson_will_2014},
\be
{\cal P} = \frac{\NewtonG}{5 c^5} \dddot{I}^{\,(jk)} \dddot{I}^{\,(jk)} , \;\;\;
I^{(jk)} = \nu m {\bf r}^j {\bf r}^k ,
\ee
where ${\cal P}$ is the rate of energy loss by gravitational waves and $I^{(jk)}$ the mass
quadrupole moment, we obtain for circular orbits
\be
{\cal P} = \frac{32 \nu^2 \NewtonG^4 m^5}{5 c^5 a^5}
\left( 1 + \frac{4\pi\rho_0 a^3}{m} \right) .
\label{eq:P-loss}
\ee
Then, the balance equation $\frac{dE}{dt} = - {\cal P}$ gives for the drift of the orbital
radius
\be
\langle \dot a \rangle_{\rm gw} = - \frac{64 \nu \NewtonG^3 m^3}{5 c^5 a^3}
\left( 1 - \frac{4\pi\rho_0 a^3}{3 m} \right) ,
\ee
which agrees with Eq.(\ref{eq:a-GW}) at $\ecc=0$ when the dark matter halo is negligible.
Although the additional halo gravity increases the radiative loss (\ref{eq:P-loss}),
this is more than compensated by the higher energy (\ref{eq:E-circular-halo}) and the orbital
drift is reduced.

\section{GW phase and the impact of dark matter}
\label{sec:phase}

\subsection{Constant mass approximation}

At lowest order, we can sum the contributions from the accretion of dark matter, the dynamical
friction and the emission of GWs. This gives the total drift of the orbital radius
\be
\langle \dot a \rangle = \langle \dot a \rangle_{\rm acc} + \langle \dot a \rangle_{\rm df}
+\langle \dot a \rangle_{\rm gw} .
\label{eq:drift-a}
\ee
This drift depends on the masses of the two BHs and their accretion rates.
However, for small accretion rates we can
take $m_i$ and $\dot m_i$ to be constant over the duration of the measurement.
Assuming this spans ${\cal N}$ orbital periods, with typically ${\cal N} \sim 100$,
we require that $\dot m_i {\cal N} P \ll m_i$.
For the maximum accretion rate (\ref{eq:accretion-rate}) this gives
\be
\rho_a \ll \frac{c^3 f}{24 \pi F_\star \NewtonG ^2 m_> {\cal N}} ,
\label{eq:rhoa-upper-acc}
\ee
where $f=2/P_{\rm orb}$ is the GW frequency (which is twice the orbital
frequency) and $m_> = \max(m_1,m_2)$.
This gives
\be
\rho_a \ll 6 \times 10^{10} \, {\cal N}^{-1} \left( \frac{m_>}{1 M_\odot} \right)^{-1} \left( \frac{f}{1 \, {\rm Hz}} \right) {\rm g} \cdot {\rm cm}^{-3} .
\label{eq:rhoa-upper-acc-1}
\ee
The strongest limitation is associated with the case of Massive Binary Black Holes (MBBH)
to be detected with the space interferometer LISA, at frequencies $f \gtrsim 10^{-4} {\rm Hz}$.
This gives the upper bound $\rho_a \ll 0.01 \, {\rm g/cm}^3$, which is much beyond the
expected dark matter densities. For instance, the dark matter density in the Solar System
is about $10^{-24} {\rm g/cm}^3$ \cite{Catena:2009mf, Weber:2009pt, Salucci:2010qr, Bovy:2012tw, Pato:2015dua, deSalas:2019pee, Lin:2019yux, Cautun:2019eaf, Sofue:2020rnl}.
On the other hand, accretion disks around supermassive BHs can have baryonic densities up to $10^{-9} {\rm g/cm^3}$ for thick disks and $10^{-1} {\rm g/cm^3}$ for thin disks \cite{Barausse:2014tra}.
Therefore, the bound (\ref{eq:rhoa-upper-acc-1})
is well satisfied up to the baryonic densities found in accretion disks.
At higher densities, we should explicitly take into account the time dependence of
the BH masses and accretion rates. This would further enhance the deviation from
the signal associated with the binary system in vacuum and increase the dark matter
impact on the waveform. Therefore, our computation provides a conservative estimate of the
detection threshold.

\subsection{Phase and coalescence time}
\label{sec:phase_time}

In the limit of small eccentricity, $\ecc \ll 1$, the drift (\ref{eq:drift-a}) reads
\ba
&& \dot{a} = -\frac{64\nu c}{5}\left(\frac{\NewtonG m}{c^2 a}\right)^3
\left( 1 - \frac{4\pi\rho_0 a^3}{3 m} \right) - a A_{\rm acc} \nonumber \\
&& - a \left( \frac{a}{\NewtonG m}\right)^{3/2} \left[ B_{\rm acc} + B_{\rm df}
+ C_{\rm df} \ln\left( \sqrt{\frac{\NewtonG m}{a}} \frac{1}{c_s} \right) \right] .
\label{eq:dot-a-tot}
\ea
The frequency $\ff$ of the gravitational waves is twice the orbital frequency
(\ref{eq:f-orb-DM}),
\be
\ff = \frac{1}{\pi} \sqrt{\frac{\NewtonG m}{a^3}}
\left( 1 + \frac{2\pi\rho_0 a^3}{3 m} \right) .
\label{eq:ff-DM}
\ee
We use a gothic font in this section to distinguish $\ff$, the function of time describing
the frequency sweep, from $f$, the Fourier-transform variable used below in the Fourier-space
analysis of the time-sequence data.
This also gives, at first order in dark matter perturbations,
\be
\dot\ff = \frac{1}{\pi} \sqrt{\frac{\NewtonG m}{a^3}} \left( \frac{\dot m}{2m}
- \frac{3 \dot a}{2 a} \right) + \NewtonG \rho_0 \left( \frac{a^3}{\NewtonG m} \right)^{1/2}
\frac{\dot a}{a} .
\ee
Together with Eqs.(\ref{eq:dot-a-tot})-(\ref{eq:ff-DM}), and using Eqs.(\ref{eq:accretion-rate})
and (\ref{eq:A-acc-B-acc-def}) to combine the accretion terms, we obtain
\be
\frac{\dot \ff}{\ff} = D_{\rm gw} + D_{\rm halo} + D_{\rm acc} + D_{\rm df}  ,
\label{eq:dot-f-f}
\ee
with
\ba
D_{\rm gw} & = & \ff^{\,8/3} \, \frac{96 \pi^{8/3} \nu}{5 c^5} (\NewtonG  m)^{5/3} , \nonumber \\
D_{\rm halo} & = & - \, \ff^{\,2/3} \, \frac{256 \pi^{5/3} \nu \rho_0 \NewtonG^{8/3} m^{5/3}}
{3 c^5} , \nonumber \\
D_{\rm acc} & = & \frac{12\pi F_\star \NewtonG^2 \rho_0 \mu}{c_s^2 c}
\sum_{i=1}^2 \Theta(\ff < f_{{\rm acc},i}) \left( 3 + 2 \frac{m_i^2}{m \mu} \right)
\nonumber \\
&& + \, \ff^{\,-1} \, 4 \NewtonG \rho_0 \sum_{i=1}^2
\Theta(\ff > f_{{\rm acc},i}) \frac{m_i^3}{\mu^2 m} \left( 3 + 2 \frac{m_i^2}{m \mu} \right) ,
\nonumber \\
D_{\rm df} & = & - \ff^{\,-1} \, \frac{4 \NewtonG \rho_0}{3} \sum_{i=1}^2
\Theta(f_{{\rm df},i}^- < \ff < f_{{\rm df},i}^+) \frac{m_i^3}{\mu^2 m}
\ln\left(\frac{\ff}{f_{{\rm df},i}^+}\right) , \nonumber \\
&& \label{eq:D-all}
\ea
and
\be
f_{{\rm acc},i} = \frac{c_s^2 c m_i^3}{3\pi F_\star \NewtonG m \mu^3} , \;\;\;
f_{{\rm df},i}^- = \frac{c_s^3 m_i^3}{\pi \NewtonG m \mu^3} , \;\;\;
f_{{\rm df},i}^+ = \frac{e^3 c_s^3 m^5 \mu^{15}}{5832 \pi \NewtonG m_i^{21}} .
\label{eq:f_acc-f_df}
\ee
In (\ref{eq:dot-f-f}) we split the contributions from gravitational waves in the
standard $\ff^{\,8/3}$ term associated with Keplerian orbits and the correction
in $\ff^{\,2/3}$ due to the dark matter halo.
Integrating the phase $\Phi(t)= 2\pi \int d\ff \, (\ff/\dot \ff)$ and the time
$t = \int d\ff \, (1/\dot \ff)$ over the GW frequency \cite{Poisson:1995ef}, we obtain
\be
\Phi(\ff) = \Phi_c + \Phi_{\rm gw} + \Phi_{\rm halo} + \Phi_{\rm acc} + \Phi_{\rm df}
\label{eq:Phi-f}
\ee
and
\be
t(\ff) = t_c + t_{\rm gw} + t_{\rm halo} + t_{\rm acc} + t_{\rm df} ,
\label{eq:t-f}
\ee
where $\Phi_c$ and $t_c$ are the phase and the time at coalescence time,
and we introduced
\ba
&& \Phi_{\rm gw} = - 2 \pi \int_{\ff}^{\infty} d\ff \frac{1}{D_{\rm gw}} , \;\;
\Phi_{\rm halo} = 2 \pi \int_{\ff}^{\infty} d\ff \frac{D_{\rm halo}}{D_{\rm gw}^2} , \dots
\nonumber \\
&& t_{\rm gw} = - \int_{\ff}^{\infty} \frac{d\ff}{\ff} \frac{1}{D_{\rm gw}} , \;\;
t_{\rm halo} = \int_{\ff}^{\infty} \frac{d\ff}{\ff} \frac{D_{\rm halo}}{D_{\rm gw}^2} , \dots
\ea
Equations (\ref{eq:Phi-f})-(\ref{eq:t-f}) provide an implicit expression for the function
$\Phi(t)$, describing the GWs phase as a function of time.
Here, we linearized over the dark matter contributions to the frequency drift, assuming
they are weaker than the Keplerian GW contribution.
As seen in Sec.~\ref{sec:relative-impact} below,
this is the case in realistic configurations. Besides, this is sufficient for the purpose
of estimating the dark matter density thresholds required for detection. At much higher
densities, our computation of the frequency drift is no longer reliable but the presence of
dark matter would remain clear in the data.

We recover the fact that the dark matter contributions are more important during the early stages
of the inspiral, that is, at low frequencies.
This means that relativistic corrections to the orbital motion would not change our results for the
dark matter detection thresholds.

The GW signal is of the form $h(t) = {\cal A}(t) \cos[\Phi(t)]$,
where $\Phi(t)$ is implicitly determined by Eqs.(\ref{eq:Phi-f})-(\ref{eq:t-f})
and ${\cal A}(t) \propto \ff^{\, 2/3}$ if we neglect the dark matter corrections in the
amplitude \cite{poisson_will_2014}.
The Fourier-space data analysis considers the Fourier transform
$\tilde h(f) = \int dt \, e^{i2\pi f t} h(t) $.
In the stationary phase approximation \cite{Poisson:1995ef}, one obtains
$\tilde h(f) = {\cal A}(f) e^{i\Psi(f)}$,
with
\be
{\cal A}(f) \propto f^{-7/6} , \;\;\;  \Psi(f) = 2\pi f t_\star - \Phi(t_\star) - \pi/4 ,
\label{eq:A-f-Psi-f}
\ee
where the saddle-point $t_\star$ is defined by $\ff(t_\star) = f$, as
$\dot \Phi= 2\pi \ff$.
Using Eqs.(\ref{eq:Phi-f})-(\ref{eq:t-f}) we obtain
\be
\Psi(f) = 2\pi f t_c - \Phi_c - \frac{\pi}{4} + \Psi_{\rm gw} + \Psi_{\rm halo}
+ \Psi_{\rm acc} + \Psi_{\rm df} ,
\label{eq:Psi-f-sum}
\ee
where the different contributions are
\ba
&& \Psi_{\rm gw} = 2 \pi \left[ \int_f^{\infty} df \frac{1}{D_{\rm gw}}
- f \int_f^{\infty} \frac{df}{f} \frac{1}{D_{\rm gw}} \right] , \nonumber \\
&& \Psi_{\rm halo} = 2 \pi \left[ f \int_f^{\infty} \frac{df}{f}
\frac{D_{\rm halo}}{D_{\rm gw}^2} - \int_f^{\infty} df \frac{D_{\rm halo}}{D_{\rm gw}^2}
\right] , \dots \hspace{0.6cm}
\label{eq:Psi-halo-int}
\ea
This gives \citep{Poisson:1995ef}
\ba
\Psi_{\rm gw} & = & \frac{3}{128} \left( \frac{\pi \NewtonG {\cal M} f}{c^3} \right)^{-5/3}
\left[ 1 + \frac{20}{9} \left( \frac{743}{336} + \frac{11}{4} \nu \right) \right.
\nonumber \\
&& \times \left. \left( \frac{\pi \NewtonG m f}{c^3} \right)^{2/3} \right] ,
\label{eq:Psi-GW}
\ea
where ${\cal M}$ is the chirp mass,
\be
{\cal M} = \nu^{3/5} m ,
\label{eq:cM-def}
\ee
and
\be
\Psi_{\rm halo} = \frac{25 \pi}{924} \frac{\rho_0 \NewtonG^3 {\cal M}^2}{c^6}
( \pi \NewtonG  {\cal M} f/c^3)^{-11/3} ,
\label{eq:Psi-halo}
\ee
\ba
&& \Psi_{\rm acc} = - \frac{25 \pi \NewtonG^3 {\cal M}^2 \rho_0}{38912 c^6}
\left( \frac{\pi \NewtonG {\cal M} f}{c^3}\right)^{-16/3} \sum_{i=1}^2 \Theta(f>f_{{\rm acc},i}) \nonumber \\
&& \times \frac{m_i^3}{\mu^2 m} \left( 3 + 2 \frac{m_i^2}{m \mu} \right)
- \frac{75 \pi F_\star \nu^{2/5} \NewtonG^3 {\cal M}^2 \rho_a}{26624 c^6}
\left( \frac{\pi \NewtonG {\cal M} f}{c^3}\right)^{-13/3} \nonumber \\
&& \times \sum_{i=1}^2 \Theta(f<f_{{\rm acc},i})
\left( 3 + 2 \frac{m_i^2}{m \mu} \right) \left[ 1 - \left( \frac{f}{f_{{\rm acc},i}} \right)^{13/3} \right.
\nonumber \\
&& \left. + \frac{13}{19} \left( \frac{f}{f_{{\rm acc},i}} \right)^{16/3} \right] ,
\label{eq:Psi-acc}
\ea
\ba
&& \Psi_{\rm df} = \frac{875 \pi \NewtonG^3 {\cal M}^2 \rho_0}{11829248 c^6}
\left( \frac{\pi \NewtonG {\cal M} f}{c^3}\right)^{-16/3} \sum_{i=1}^2 \frac{m_i^3}{\mu^2 m}
 \Theta(f_{{\rm df},i}^- \!<\!  f_{{\rm df},i}^+) \nonumber \\
&& \times \left\{ \Theta(f_{{\rm df},i}^- \!<\! f \!<\!  f_{{\rm df},i}^+) \left[ 1+
\frac{304}{105} \ln \frac{f}{f_{{\rm df},i}^+}  - \frac{361}{105}  \left( \frac{f}{f_{{\rm df},i}^+} \right)^{16/3} \right.
\right. \nonumber \\
&& \left. +  \frac{256}{105} \left( \frac{f}{f_{{\rm df},i}^+} \right)^{19/3} \right] + \Theta(f < f_{{\rm df},i}^-)
 \left[ - \frac{361}{105} \left( \frac{f}{f_{{\rm df},i}^+} \right)^{16/3} \right. \nonumber \\
&& + \frac{361}{105}
\left( \frac{f}{f_{{\rm df},i}^-} \right)^{16/3} + \frac{5776}{315} \left( \frac{f}{f_{{\rm df},i}^-} \right)^{16/3}
\ln \frac{f_{{\rm df},i}^-}{f_{{\rm df},i}^+}  + \frac{256}{105} \nonumber \\
&& \left. \left. \! \times \! \left( \! \frac{f}{f_{{\rm df},i}^+} \! \right)^{19/3} \!\!\!
-  \frac{256}{105} \left( \! \frac{f}{f_{{\rm df},i}^-} \! \right)^{19/3} \!\!\! - \frac{4864}{315}
\left( \! \frac{f}{f_{{\rm df},i}^-} \! \right)^{19/3} \!\!\! \ln \frac{f_{{\rm df},i}^-}{f_{{\rm df},i}^+}
\right] \right \} .  \nonumber \\
&&
\label{eq:Psi-df}
\ea
The factor $\Theta$ in the first line means that only the smaller BH can contribute, if there exists
a range for dynamical friction where the two conditions $v_i > c_s$ and $r_{{\rm IR},i} > r_{{\rm UV},i}$
are satisfied.
This provides a conservative estimate of the impact of the dark matter environment on the gravitational
wave signal. A more accurate treatment would probably give a nonzero dynamical friction outside of the
frequency ranges $[ f_{{\rm df},i}^- , f_{{\rm df},i}^+ ]$. Therefore, the detection thresholds obtained
in Table~\ref{tab:rho_a_detect} are conservative results. However, as the signal is dominated by the accretion
rather than the dynamical friction, more accurate treatments of the dynamical friction that would give
a small but non-zero impact outside of these frequency ranges should not change much our results.

In the dark matter contributions (\ref{eq:Psi-halo})-(\ref{eq:Psi-df}) to the phase
we used the leading term $D_{\rm gw}$ given in (\ref{eq:D-all}) in the expressions
(\ref{eq:Psi-halo-int}). This is sufficient for our purpose, which is to estimate
the dark matter density thresholds associated with a significant impact on the GW signal.
However, in the gravitational wave phase (\ref{eq:Psi-GW}) we have added the first
post-Newtonian 1-PN order \citep{Poisson:1995ef}. This breaks the degeneracy over the two BH
masses $m_1$ and $m_2$ shown by the leading term that only depends on the chirp mass ${\cal M}$.
Then, the phase (\ref{eq:Psi-GW}) depends independently on both $m_1$ and $m_2$ and the
gravitational wave signal can constrain both BH masses.
Higher-order 1.5-PN and 2-PN terms allow one to constrain the BH spins \citep{Poisson:1995ef},
however we do not consider BH spins in this paper.
This ensures that for vanishing dark matter density, i.e. a binary in vacuum, the Fisher
analysis performed in Sec.~\ref{sec:fish} over the binary parameters
$\{t_c, \Phi_c, \ln(m_1), \ln(m_2) \}$ is well defined and can constrain both BH masses,
as in actual data analysis of GW signals.

\subsection{Relative impact of various contributions}
\label{sec:relative-impact}

\subsubsection{Dark matter halo gravity}

From Eqs.(\ref{eq:Psi-GW}) and (\ref{eq:Psi-halo}), we obtain
\be
\frac{\Psi_{\rm halo}}{\Psi_{\rm gw}} = \frac{800 \rho_0 \NewtonG}{693 \pi f^2}  \sim 2 \times 10^{-8}
\frac{\rho_0}{1 \, {\rm g}\cdot {\rm cm}^{-3}} \left(\frac{f}{1 \, {\rm Hz}}\right)^{-2} ,
\label{eq:cloud-gravity}
\ee
where we only kept the leading term in $\Psi_{\rm gw}$.
This ratio happens to be independent of the BH masses and is very small.
Therefore, the impact of the dark matter cloud gravitational potential is typically negligible.

\subsubsection{Accretion on the BHs}
\label{sec:accretion-ratio}

Denoting $m_>=\max(m_1,m_2)$ and $m_<=\min(m_1,m_2)$ the greater and smaller mass of the
binary, we obtain from Eq.(\ref{eq:f_acc-f_df})
\ba
&& f_{{\rm acc},<} \sim 3 \times 10^4 \frac{\rho_0}{\rho_a} \left( \frac{m_>}{1 \, M_\odot}\right)^{-1} {\rm Hz} ,
\nonumber \\
&& f_{{\rm acc},>} \sim 3 \times 10^4 \frac{\rho_0}{\rho_a} \left( \frac{m_>}{m_<} \right)^3
\left( \frac{m_>}{1 \, M_\odot}\right)^{-1} {\rm Hz} .
\ea
Since we typically have $\rho_0 \ll \rho_a$, these frequencies are usually below $1$~Hz and the smaller
BH can experience both accretion regimes in the range of frequencies probed by observations.
The impact of the accretion is typically greater for the more massive BH, because of the factors
$m_i^3$ and $m_i^2$ in Eq.(\ref{eq:Psi-acc}). Focusing on this contribution, we obtain
\ba
f > f_{{\rm acc},>} : && \frac{\Psi_{{\rm acc},>}}{\Psi_{\rm gw}} \sim 0.1
\left( \frac{m_>}{m_<} \right)^4 \frac{\rho_0}{1 \, {\rm g}\cdot {\rm cm}^{-3}}  \nonumber \\
&& \times \left( \frac{m_>}{1 \, M_\odot}\right)^{-5/3} \left(\frac{f}{1 \, {\rm Hz}}\right)^{-11/3} ,
\label{eq:ratio-acc}
\ea
and
\ba
f < f_{{\rm acc},>} : && \frac{\Psi_{{\rm acc},>}}{\Psi_{\rm gw}} \sim 5 \times 10^{-6}
\; \frac{m_>}{m_<} \; \frac{\rho_a}{1 \, {\rm g}\cdot {\rm cm}^{-3}} \nonumber \\
&& \times \left( \frac{m_>}{1 \, M_\odot}\right)^{-2/3} \left(\frac{f}{1 \, {\rm Hz}}\right)^{-8/3} .
\ea
We can see that the contribution to the phase from the accretion is typically much greater than that
from the cloud gravity (\ref{eq:cloud-gravity}).
However, it remains small as compared with the standard contribution $\Psi_{\rm gw}$ from gravitational
waves, which validates our perturbative computations.
It increases for smaller masses and low frequencies. This implies that it is most important at the early stages
of the inspiral phase.

\subsubsection{Dynamical friction}

From Eq.(\ref{eq:f_acc-f_df}) we obtain
\ba
&& f_{{\rm df},<}^- \sim 6 \times 10^4 \left( \frac{\rho_0}{\rho_a} \right)^{3/2}
\left( \frac{m_>}{1 \, M_\odot}\right)^{-1} {\rm Hz}  , \nonumber \\
&&  f_{{\rm df},>}^- \sim 6 \times 10^4 \left( \frac{\rho_0}{\rho_a} \right)^{3/2}
\left( \frac{m_>}{m_<} \right)^3 \left( \frac{m_>}{1 \, M_\odot}\right)^{-1} {\rm Hz} , \hspace{1cm}
\ea
and
\ba
&& f_{{\rm df},<}^+ \sim 2 \times 10^2 \left( \frac{\rho_0}{\rho_a} \right)^{3/2}
\left( \frac{m_>}{m_<} \right)^6 \left( \frac{m_>}{1 \, M_\odot}\right)^{-1} {\rm Hz}  , \nonumber \\
&&  f_{{\rm df},>}^+ \sim 2 \times 10^2 \left( \frac{\rho_0}{\rho_a} \right)^{3/2}
\left( \frac{m_<}{m_>} \right)^{15} \left( \frac{m_>}{1 \, M_\odot}\right)^{-1} {\rm Hz} . \hspace{1cm}
\ea
We recover the fact that only the smaller BH experiences a significant dynamical friction, if the
mass ratio is sufficiently large.
Then, we obtain
\ba
f_{{\rm df},<}^- < f < f_{{\rm df},<}^+ : && \frac{\Psi_{\rm df}}{\Psi_{\rm gw}} \sim 7 \times 10^{-3}
\frac{\rho_0}{1 \, {\rm g}\cdot {\rm cm}^{-3}} \nonumber \\
&& \times \left( \frac{m_>}{1 \, M_\odot}\right)^{-5/3} \left(\frac{f}{1 \, {\rm Hz}}\right)^{-11/3} .
\hspace{0.6cm}
\ea
This is smaller than the accretion contribution (\ref{eq:ratio-acc}) by a factor $(m_</m_>)^4$
because the accretion is dominated by the larger BH while in our approximation only the smaller BH
experiences dynamical friction.
Again this is a small correction to the gravitational wave term $\Psi_{\rm gw}$ from gravitational
waves and it is most important at the early stages of the inspiral phase, with low frequencies.

\subsubsection{Effective post-Newtonian orders}

Contributions to the phase $\Psi$ that scale as $f^{\alpha}$ may be attributed an effective
post-Newtonian order $n=3\alpha/2+5/2$. Then, the cloud gravity (\ref{eq:Psi-halo})
is associated with a -3~PN contribution. The accretion gives a -4~PN contribution at low
frequency and a -5.5~PN contribution at high frequency, keeping only the dominant terms.
In the range $f_{\rm df}^-<f<f_{\rm df}^+$ the dynamical friction also gives
a -5.5~PN contribution.
This negative orders express the fact that these dark matter contributions are increasingly
important at low frequencies, in the early stages of the inspiral.
This also means that they are not degenerate with usual relativistic corrections, associated
with positive post-Newtonian orders.

In this paper we do not include the backreaction of the scalar field. Studies of the FDM scenario
have shown that this may contribute a -6~PN effect, which is however too small to be observed
\cite{Annulli:2020lyc, Annulli:2020ilw}.
On the other hand, the dynamical friction can heat the gas and lead to a depletion of dark matter
in the vicinity of the orbital radius \cite{Kavanagh:2020cfn, Kim:2022mdj}, which decreases the
actual amount of dynamical friction.
For the self-interacting case that we consider in this paper, the effective pressure could lessen this
effect if it can replenish the BH neighbourhood. Moreover, the small-scale cutoff (\ref{eq:r-UV})
makes the dynamical friction insensitive to the local dark matter density.
A detailed investigation of this point is left to future work.
Another noteworthy factor, at 5~PN order, is the influence of deformability effects caused by
nonzero Love numbers for dressed BHs (e.g., surrounded by a scalar field) as discussed in
\cite{DeLuca:2021ite, DeLuca:2022xlz}.
Focusing on the low scalar-mass limit for FDM models,
$\alpha = m_{\rm DM} m_{\rm BH} G/(\hbar c) \ll 1$, these authors
found that these effects grow as $\alpha^{-8}$ and can be significant for $\alpha \lesssim 0.1$.
In this paper, we focus instead on the large scalar-mass limit, $\alpha \gg 1$ as in Eq.(\ref{eq:large-m-rs}),
and we can expect the tidal Love numbers to be much smaller. Another difference is the importance
of the self-interactions. We plan to study the Love numbers in this case in future papers.

\subsubsection{Relativistic corrections}
\label{sec:non-rel}

The dynamical friction formulae used here are valid in the nonrelativist limit $v \ll c$.
Relativistic corrections typically give a corrective prefactor
$\gamma^2 (1+v^2)^2$ in the dynamical friction
\cite{Syer:1994vr,Barausse:2007ph,Traykova:2021dua},
which enhances the impact on the binary and the detectability of the environment
\citep{Speeney_2022}.
This can be obtained in the collisionless case from the relativistic formula for the
scattering deflection angle and the relativistic Lorentz boost between the fluid and
BH frames \cite{Syer:1994vr}.
The relativistic corrections for fuzzy dark matter were also derived from first principles
in \cite{Vicente:2022ivh} and compared with numerical simulations in \cite{Traykova:2023qyv}.
This should remain a good approximation in the highly
supersonic case, where the streamlines at large radii follow collisionless trajectories
as pressure effects are small.
For velocities as high as $v^2 \sim 0.137 \, c^2$ this only gives a multiplicative factor
of about $1.5$. As the dark matter contributions are most important in the early inspiral,
we can see that relativistic corrections can be neglected and will not change the
order of magnitude of our results. In practice, we cut the analysis below the
frequency $f_{\gamma}$ where $v^2=0.137 \, c^2$, to ensure relativistic corrections
remain modest.

\subsubsection{Dark matter parameters $\rho_a$ and $\rho_0$}

As seen in the previous sections, the gravitational wave signal only depends on the dark
matter environment through the two parameters $\rho_a$ and $\rho_0$, which are the
characteristic density (\ref{eq:ra-def}) determined by the self-interaction and the bulk
density of the dark matter cloud.
The cloud gravity (\ref{eq:Psi-halo}), the accretion at high frequency (\ref{eq:Psi-acc})
and the dynamical friction (\ref{eq:Psi-df}) are proportional to $\rho_0$,
whereas the accretion at low frequency (\ref{eq:Psi-acc}) is proportional to $\rho_a$.
On the other hand, the thresholds (\ref{eq:f_acc-f_df}) depend on
$c_s \propto \sqrt{\rho_0/\rho_a}$.
Therefore, in principles it is possible to constrain both parameters if the observational
frequency range contains the low-frequency accretion regime or at least one of these
frequency thresholds.

\section{Fisher Information Matrix}
\label{sec:fish}

\subsection{Fisher analysis}

We use a Fisher matrix analysis to estimate the dark matter densities $\rho_a$ and
$\rho_0$ that could be detected through the measurement of GWs emitted
by binary BHs in the inspiral phase.
The Fisher matrix is given by \cite{Poisson:1995ef,Vallisneri:2007ev}
\be
\Gamma_{ij} = 4 \, {\rm Re} \int_{f_{\min}}^{f_{\max}} \frac{df}{S_n(f)} \,
\left( \frac{\partial\tilde h}{\partial\theta_i} \right)^\star
\left( \frac{\partial\tilde h}{\partial\theta_j}\right) ,
\ee
where $\{\theta_i\}$ is the set of parameters that we wish to measure and
$S_n(f)$ is the noise spectral density, which depends
on the GW interferometer.
The signal-to-noise ratio is
\be
({\rm SNR})^2 = 4 \int_{f_{\min}}^{f_{\max}} \frac{df}{S_n(f)} \, | \tilde h(f) |^2 .
\ee
Writing the gravitational waveform as $\tilde{h}(f) = {\cal A}_0 f^{-7/6} e^{i\Psi(f)}$,
as in Eqs.(\ref{eq:A-f-Psi-f})-(\ref{eq:Psi-f-sum}),
we obtain
\be
\Gamma_{ij} = \frac{({\rm SNR})^2}{\int_{f_{\min}}^{f_{\max}} \frac{df}{S_n(f)} f^{-7/3}}
\int_{f_{\min}}^{f_{\max}} \frac{df}{S_n(f)} f^{-7/3} \frac{\partial\Psi}{\partial\theta_i}
\frac{\partial\Psi}{\partial\theta_j}
\label{eq:Fisher-def}
\ee
where the parameters that we consider in our analysis are
$\{\theta_i\} = \{t_c, \Phi_c, \ln(m_1), \ln(m_2), \rho_0, \rho_a \}$.
The amplitude ${\cal A}_0$ would be an additional parameter. However, the Fisher matrix
is block-diagonal as $\Gamma_{{\cal A}_0,\theta_i}=0$ and the amplitude ${\cal A}_0$
is completely decorrelated from the other parameters $\{\theta_i\}$ \cite{Poisson:1995ef}.
Therefore, we do not consider the amplitude any further.
From the Fisher matrix we obtain the covariance
$\Sigma_{ij} = \left(\Gamma^{-1}\right)_{ij}$, which gives the standard deviation on the
various parameters as
$\sigma_i = \langle (\Delta\theta_i)^2 \rangle^{1/2} = \sqrt{\Sigma_{ii}}$.

As compared with the study presented in \cite{Cardoso:2019rou}, we neglect the effective
spin $\chi_{\rm eff} \equiv (m_1\chi_1 + m_2\chi_2)/m$, which is only considered  to calculate the last stable orbit using the analytical PhenomB templates \cite{PhysRevLett.106.241101}.
This is because our results for the accretion rate and the dynamical friction have only been
derived for Schwarzschild BHs. However, we expect the order of magnitude that we obtain
for the dark matter densities to remain valid for moderate spins.
A second difference from \cite{Cardoso:2019rou} is that in addition to the dark-matter density
$\rho_0$, which describes the bulk of the cloud, we also have a second characteristic
density $\rho_a$. It describes the dark matter density close to the
Schwarzschild radius and it is directly related to the strength of the dark-matter
self-interaction.

\subsection{Sectors in the $(\rho_0,\rho_a)$ plane}

\subsubsection{Binary and dark matter parameters}

In this paper, we investigate the detection thresholds for a dark matter environment.
Then, we assumed that the dark matter impact is small and we linearized in all its
contributions.
Thus, the phases (\ref{eq:Psi-halo})-(\ref{eq:Psi-df}) are proportional
to the densities $\rho_0$ or $\rho_a$ (at fixed $c_s$).
As expected, the contributions from the halo gravity (\ref{eq:Psi-halo}), the accretion in the
high-frequency or high-velocity regime (\ref{eq:Psi-acc}), and the dynamical friction
(\ref{eq:Psi-df}) are proportional to the bulk halo density $\rho_0$.
The contribution from the accretion in the low-frequency or low-velocity regime
(\ref{eq:Psi-acc}) is proportional to the characteristic density $\rho_a$,
associated with the maximum allowed accretion rate.

Then, for vanishing or negligible dark matter halo the standard waveform parameters
$\{\theta_i\}_{i=1,4} = \{t_c, \Phi_c, \ln(m_1), \ln(m_2) \}$ are determined by the
first four terms in the phase (\ref{eq:Psi-f-sum}), that is, the $t_c$ and $\Phi_c$ factors
and the gravitational wave contribution $\Psi_{\rm gw}$.
This corresponds to the standard analysis for binary systems in vacuum.
For a small dark matter halo, or for the fiducial $\rho_0=\rho_a=0$, this also provides the
$4\times 4$ components $\Gamma_{ij}$ with $1 \leq i,j \leq 4$ of the Fisher matrix.

The presence of a dark matter environment can be detected through the phases
(\ref{eq:Psi-halo})-(\ref{eq:Psi-df}).
These contributions have an amplitude proportional to $\rho_0$ or $\rho_a$, multiplied
Heaviside factors $\Theta$ and slowly-varying terms such as $1+(f/f_{\rm acc})^{13/3}$
or $\ln(f/f_{\rm df}^+)$.
The frequencies (\ref{eq:f_acc-f_df}) do not depend on $\rho_0$ and $\rho_a$ independently,
but only on the sound-speed $c_s$, that is, on the ratio $y$ defined by
\be
y \equiv \frac{\rho_a}{\rho_0} = \frac{c^2}{c_s^2} \geq 1 .
\label{eq:y-def}
\ee
Therefore, the different accretion and dynamical friction regimes are delimited by
specific values of $y$, which determine several angular sectors in the $(\rho_0,\rho_a)$ plane.
The physical part of the positive quadrant $\{\rho_0 \geq 0, \rho_a \geq 0\}$ is restricted
to the upper-diagonal sector $\rho_a\geq \rho_0$ because of the condition $c_s \leq c$.
For a given binary system and observational frequency band
$[f_{\min},f_{\max}]$, let us define the accretion thresholds in $y$,
\be
f_{\rm min} < f_{{\rm acc},i} : \;\; y < y_{{\rm acc},i}^+ , \;\;
y_{{\rm acc},i}^+ = \frac{c^3 m_i^3}{3 \pi F_\star \NewtonG m \mu^3 f_{\rm min}} ,
\ee
\be
f_{\rm max} < f_{{\rm acc},i} : \;\; y < y_{{\rm acc},i}^- , \;\;
y_{{\rm acc},i}^- = \frac{c^3 m_i^3}{3 \pi F_\star \NewtonG m \mu^3 f_{\rm max}} ,
\ee
and the dynamical friction thresholds
\be
f_{\rm min} < f_{{\rm df},i}^+ : \;\; y < y_{{\rm df},i}^+ , \;\;
y_{{\rm df},i}^+ = \left( \frac{e^3 c^3 m^5 \mu^{15}}{5832 \pi \NewtonG m_i^{21} f_{\rm min}}
\right)^{2/3} ,
\ee
\be
f_{\rm max} < f_{{\rm df},i}^- : \;\; y < y_{{\rm df},i}^- , \;\;
y_{{\rm df},i}^- = \left( \frac{c^3 m_i^3}{\pi \NewtonG m \mu^3 f_{\rm max}} \right)^{2/3} .
\ee
Let us label the BH masses so that $m_1 \geq m_2$, then we have
\be
m_1 \geq m_2 : \;\;  y_{{\rm acc},1}^+ \geq y_{{\rm acc},2}^+ , \;\;
y_{{\rm acc},1}^- \geq y_{{\rm acc},2}^- ,
\ee
while only the smaller BH $m_2$ can experience significant dynamical friction.
Then, we can split the behavior of the accretion term $\Psi_{\rm acc}$ as
\ba
&& y > y_{{\rm acc},1}^+ : \;\; \mbox{no accretion dependence on} \;\; \rho_a , \nonumber \\
&& y < y_{{\rm acc},2}^- : \;\; \mbox{no accretion dependence on} \;\; \rho_0 ,
\ea
where we neglected the dependence on $c_s$ of the terms inside the brackets
in Eq.(\ref{eq:Psi-acc}), which quickly converge to unity below the threshold
$f_{{\rm acc},i}$.
We can also split the behavior of the dynamical friction term $\Psi_{\rm df}$ as
\ba
&& y > y_{{\rm df},2}^+ : \;\; \mbox{no dynamical friction} , \nonumber \\
&& y_{{\rm df},2}^- < y < y_{{\rm df},2}^+ : \;\; \mbox{dynamical friction}  , \nonumber \\
&& y < y_{{\rm df},2}^- : \;\; \mbox{dynamical friction is degenerate with $t_c$ and $\Phi_c$} ,
\nonumber \\
\ea
where again we neglected the dependence on $c_s$ of the terms inside the brackets
in Eq.(\ref{eq:Psi-df}).

\subsubsection{High-$y$ sector}
\label{sec:high-y}

In the high-$y$ sector,
\be
y > \max(y_{{\rm acc},1}^+ , y_{{\rm df},2}^+ ) ,
\label{eq:large-y}
\ee
the phase $\Psi$ is only sensitive to $\rho_0$, through the halo gravity (\ref{eq:Psi-halo})
and the high-frequency regime of the accretion (\ref{eq:Psi-acc}).
Therefore, we have no constraint on $\rho_a$ and the gravitational wave measurement
only provides a bound on the bulk density $\rho_0$.
The Fisher matrix (\ref{eq:Fisher-def}) is then a $5\times 5$ matrix.
This gives the covariance matrix $\Sigma_{ij} = \left(\Gamma^{-1}\right)_{ij}$
and the standard deviation $\sigma_{\rho_0} = \sqrt{\Sigma_{\rho_0 \rho_0}}$.
This corresponds to the detection threshold $\rho_{0\star} = \sigma_{\rho_0}$:
halos with a higher dark matter density can be detected from the gravitational
wave measurements whereas lower density clouds cannot be discriminated from
binaries in vacuum.
This corresponds for instance in the EMRI panel in Fig.~\ref{fig:LISA_results}
to the vertical blue line above the upper red diagonal line, which is the
lower angular bound (\ref{eq:large-y}) in the plane $(\rho_0,\rho_a)$.

As seen in Sec.~\ref{sec:accretion-ratio}, the contribution from the halo gravity
is negligible as compared with the contribution from the accretion.
Then, in the limit where we can neglect the correlations between the binary parameters
$\{t_c, \Phi_c, \ln(m_1), \ln(m_2) \}$ and $\rho_0$, the detection threshold $\rho_{0\star}$
can be estimated as $ \rho_{0\star} \gtrsim \frac{1}{\rm SNR}
\left| \frac{\partial \Psi_{\rm acc}}{\partial\rho_0} \right|^{-1}$,
\be
\rho_{0\star} \gtrsim \frac{1}{{\rm SNR}} \frac{19456 c^6}{25 \pi \NewtonG^3 m_1^2}
\left( \frac{\pi \NewtonG m_1 f_{\rm min}}{c^3}\right)^{16/3} \left(\frac{m_2}{m_1}\right)^5 ,
\ee
which gives
\be
\rho_{0\star} \gtrsim \frac{3 \times 10^{-6}}{{\rm SNR}} \left(\frac{m_2}{m_1}\right)^5
\left(\frac{m_1}{1 \, M_\odot}\right)^{10/3}
\left( \frac{f_{\rm min}}{1 \, {\rm Hz}} \right)^{16/3} {\rm g/cm}^3 .
\label{eq:rho0-star}
\ee
Thus, we can see that this lower bound improves for instruments that probe
lower frequencies and for binaries with a higher mass ratio.
In practice, we perform a full Fisher matrix analysis. Then, the partial degeneracies
between the various parameters and the finite frequency band $[f_{\rm min},f_{\rm max}]$
give a detection threshold that must be somewhat above (\ref{eq:rho0-star}).

\subsubsection{Intermediate-$y$ sector}
\label{sec:intermediate-y}

For the IMRI and EMRI cases to be discussed in Sec.~\ref{sec:detection} below,
there is a narrow intermediate regime where dynamical friction comes into play
while accretion is still independent of $\rho_a$,
\be
y_{{\rm df},2}^- < y_{{\rm acc},1}^+ < y < y_{{\rm df},2}^+ .
\label{eq:intermediate-y}
\ee
Neglecting the dependence on $c_s$ of the terms inside the brackets
in Eq.(\ref{eq:Psi-df}) to count the number of parameters, we treat $\Psi_{\rm df}$
as a linear function of $\rho_0$ for a fixed density ratio $y$.
Then, the Fisher matrix (\ref{eq:Fisher-def}) is again a $5\times 5$ matrix
and from the standard deviation $\sigma_{\rho_0} = \sqrt{\Sigma_{\rho_0 \rho_0}}$
we again obtain the lower bound $\rho_{0\star}=\sigma_{\rho_0}$.
This provides a vertical boundary line in the plane $(\rho_0,\rho_a)$ for the detection
threshold, within the narrow strip (\ref{eq:intermediate-y}).
This corresponds for instance in the EMRI panel in Fig.~\ref{fig:LISA_results}
to the vertical dashed green line between the upper red diagonal line and the upper blue dotted diagonal line,
associated with the angular bounds (\ref{eq:intermediate-y}) in the plane $(\rho_0,\rho_a)$.

\subsubsection{Low-$y$ sector}

For low values of $y$,
\be
1 \leq y < y_{{\rm acc},1}^+ ,
\label{eq:low-y}
\ee
the accretion contribution depends on $\rho_a$, while the halo gravity always depends
on $\rho_0$, so that we have two dark matter parameters and the Fisher matrix is
a $6\times 6$ matrix. For a given density ratio $y$, we compute the associated
Fisher ellipse in the plane $(\rho_0,\rho_a)$ and its intersection with the direction
$\rho_a/\rho_0=y$.
Thus, from the $6\times 6$ Fisher matrix $\Gamma_{ij}$ we obtain the $6\times 6$
covariance matrix $\Sigma_{ij}$. We marginalize over the binary parameters
$\{t_c, \Phi_c, \ln(m_1), \ln(m_2) \}$ by defining the new $2\times 2$
covariance matrix $\hat\Sigma_{ij}$ associated with the rows and columns of the two remaining
parameters $\rho_0$ and $\rho_a$, and we obtain the $2\times 2$ Fisher matrix
$\hat\Gamma = \hat\Sigma^{-1}$.
This determines the Fisher ellipse in the plane $(\rho_0,\rho_a)$ defined by
\be
\Delta\chi^2 = \hat\Gamma_{\rho_0\rho_0} \rho_0^2 + 2 \hat\Gamma_{\rho_0\rho_a} \rho_0 \rho_a
+ \hat\Gamma_{\rho_a\rho_a} \rho_a^2 ,
\label{eq:Gamma-ellipse}
\ee
which is restricted to the angular sector (\ref{eq:low-y}) in the plane $(\rho_0,\rho_a)$.
For simplicity we keep $\Delta\chi^2=1$ as in the other angular sectors.
Because most of the dark matter signal comes from the accretion contribution at
low frequency, this elliptic section is an almost straight horizontal line in the angular sector
(\ref{eq:low-y}), which gives an almost constant threshold $\rho_a$.
This corresponds for instance in the EMRI panel in Fig.~\ref{fig:LISA_results}
to the horizontal red line between the upper blue dotted line and the black dashed line,
associated with the angular bounds (\ref{eq:low-y}) in the plane $(\rho_0,\rho_a)$.

Neglecting correlations among parameters we obtain the estimate
$ \rho_{a\star} \gtrsim \frac{1}{\rm SNR}
\left| \frac{\partial \Psi_{\rm acc}}{\partial\rho_a} \right|^{-1}$,
\be
\rho_{a\star} \gtrsim \frac{1}{{\rm SNR}} \frac{13312 c^6}{75 \pi F_\star \NewtonG^3 m_1^2}
\left( \frac{\pi \NewtonG m_1 f_{\rm min}}{c^3}\right)^{13/3}
\left(\frac{m_2}{m_1}\right)^2 ,
\ee
which gives
\be
\rho_{a\star} \gtrsim \frac{0.08}{{\rm SNR}} \left(\frac{m_2}{m_1}\right)^2
\left(\frac{m_1}{1 \, M_\odot}\right)^{7/3}
\left( \frac{f_{\rm min}}{1 \, {\rm Hz}} \right)^{13/3} {\rm g/cm}^3 .
\label{eq:rhoa-star}
\ee
This lower bound again improves for instruments that probe
lower frequencies and for binaries with a higher mass ratio.
Again, because of partial degeneracies and the finite frequency band the detection
threshold obtained from the inversion of the Fisher matrix is somewhat greater
than the estimate (\ref{eq:rhoa-star}).

\subsubsection{Detection area in the plane $(\rho_0,\rho_a)$}

As displayed for instance in the EMRI panel in Fig.~\ref{fig:LISA_results},
the thresholds $\rho_{0\star}$ obtained at large $y$ in Secs.~\ref {sec:high-y}
and \ref{sec:intermediate-y} give a degenerate Fisher ellipse that is a vertical strip
around $\rho_0=0$ of width $\rho_{0\star}$ that extends from the diagonal
$\rho_a>y_{\rm acc,1}^+ \rho_0$ to infinite $\rho_a$.
At lower $y$ the ellipse (\ref{eq:Gamma-ellipse}) gives an almost horizontal strip
around $\rho_a=0$ of width $\rho_{a\star}$, which is bracketed by the diagonals
$\rho_0 = \rho_a/y_{\rm acc,1}^+$ and $\rho_0=\rho_a$.
In Fig.~\ref{fig:LISA_results} this corresponds to the white area in the upper left diagonal
sector, $\rho_a \geq \rho_0$.
The shaded complementary area corresponds to densities that are beyond these
Fisher ellipse boundaries, that is, their dark matter impact on the gravitational waveform
is statistically inconsistent with the assumption of zero dark matter environment.
In this paper, we thus identify this region with the detection threshold for the dark matter
densities (i.e., dark matter environments that can be distinguished from the null hypothesis).
Although more sophisticated data analysis may be considered, this should provide the
correct order of magnitude for the detection thresholds in the dark matter density plane
$(\rho_0,\rho_a)$.

\section{Detection prospects}
\label{sec:detection}

\subsection{Gravitational-wave detectors}
\label{sec:detectors}

The gravitational-wave detectors that we consider are LISA \cite{LISA:2017pwj}, DECIGO \cite{Kawamura:2020pcg}, ET \cite{Punturo:2010zz} and  Adv-LIGO \cite{LIGOScientific:2014pky}.
We use the noise spectral densities presented in
\cite{Barsotti:2018advligo, Hild:2010id, LISA:2022kgy, Isoyama:2018rjb}.
The frequency ranges are given in
Table~\ref{table:det-frequ}, where the PhenomB inspiral-merger transition value $f_1$
is defined in \cite{PhysRevLett.106.241101} and
$f_{\rm obs} = 4.149\times10^{-5}\left(\frac{{\cal M}}{10^6 {M_\odot}}\right)^{-5/8}\left(\frac{T_{\rm obs}}{1 \, {\rm yr}}\right)^{-3/8}$ is the frequency at a
given observational time   before the merger, as defined in \cite{PhysRevD.71.084025}.
We take $T_{\rm obs} = 4$~yr in our computations.

\begin{table}[hbtp]
\centering
\begin{tabular}{|l||*{3}{c|}}
 \hline
 \backslashbox{Detector}{Frequency} & $f_{\rm min}({\rm Hz})$ & $f_{\rm max}({\rm Hz})$ \\ [0.5ex]
 \hline\hline
 \rule{0pt}{10pt} LISA  & ${\rm max}\left(2\times10^{-5}, f_{\rm obs}\right)$ & ${\rm min}\left(1, f_{\rm 1}, f_\gamma \right)$ \\
 \hline
 \rule{0pt}{10pt} DECIGO  & $10^{-2}$ & ${\rm min}\left(100, f_{\rm 1}, f_\gamma \right)$ \\
 \hline
 \rule{0pt}{10pt} ET  & $3$ & ${\rm min}\left(f_{\rm 1}, f_\gamma \right)$ \\
 \hline
 \rule{0pt}{10pt} Adv-LIGO  & $10$ & ${\rm min}\left(f_{\rm 1}, f_\gamma \right)$\\
 \hline
\end{tabular}
\caption{Gravitational waves frequency band considered for the LISA, DECIGO, ET and Adv-LIGO interferometers, where $f_{\rm obs}$ is the frequency of the binary 4 years before the merger \cite{PhysRevD.71.084025} and $f_{\rm 1}$ is the PhenomB inspiral-merger transition value \cite{PhysRevLett.106.241101}.}
\label{table:det-frequ}
\end{table}

\subsection{Events}

We focus on the description of 6 events, 2 ground based and 4 space based, the last ones being for LISA since its detection range differs from the others. All the events are BH binaries. The virtual events correspond to different types of binaries: Massive Binary Black Holes (MBBH), Intermediate Binary Black Holes (IBBH), an Intermediate Mass Ratio Inspiral (IMRI) and an Extreme Mass Ratio Inspiral (EMRI).
All of these events are of the same type as the ones considered by \cite{Cardoso:2019rou}, but
we focus on BH binaries and do not consider neutron star binaries.
The details of these events are given in Table~\ref{table:events}.
For completeness, we included the spins and $\chi_{\rm eff}$, which sets the upper
frequency cutoff of the data analysis.
The SNR values for each of these events are taken from \cite{Cardoso:2019rou}
and summarized in Table~\ref{table:SNR-events}.

\begin{table}[hbtp]
\centering
\begin{tabular}{|l||*{5}{c|}}
 \hline
 \backslashbox{Event}{Properties} & $m_1$ ($\rm M_{\odot}$) & $m_2$ ($\rm M_{\odot}$) & $\chi_1$ & $\chi_2$ & $\chi_{\rm eff}$\\ [0.5ex]
 \hline\hline
 \rule{0pt}{10pt} MBBH & $10^6$ & $5\times10^5$ & $0.9$ & $0.8$ & $0.87$ \\
 \hline
 \rule{0pt}{10pt} IBBH & $10^4$ & $5\times10^3$ & $0.3$ & $0.4$ & $0.33$ \\
 \hline
 \rule{0pt}{10pt} IMRI & $10^4$ & $10$ & $0.8$ & $0.5$ & $0.80$ \\
 \hline
 \rule{0pt}{10pt} EMRI & $10^5$ & $10$ & $0.8$ & $0.5$ & $0.80$ \\
 \hline
 \rule{0pt}{10pt} GW150914 & $35.6$ & $30.6$ &  &  & $-0.01$\\
 \hline
 \rule{0pt}{10pt} GW170608 & $11$ & $7.6$ &  &  & $0.03$\\
 \hline
\end{tabular}
\caption{Details on masses and spins of the considered events. The information on GW150914 and GW170608 are taken from \cite{LIGOScientific:2018mvr}.}
\label{table:events}
\end{table}

\begin{table}[hbtp]
\centering
\begin{tabular}{|l||*{5}{c|}}
 \hline
 \backslashbox{Event}{Detector} & LISA & DECIGO & ET & Adv-LIGO\\ [0.5ex]
 \hline\hline
 \rule{0pt}{10pt} MBBH & $3\times 10^4$ & $\times$ & $\times$ & $\times$  \\
 \hline
 \rule{0pt}{10pt} IBBH & $708$ & $\times$ & $\times$ & $\times$ \\
 \hline
 \rule{0pt}{10pt} IMRI & $64$ & $\times$ & $\times$ &  $\times$ \\
 \hline
 \rule{0pt}{10pt} EMRI & $22$ & $\times$ & $\times$ & $\times$ \\
 \hline
 \rule{0pt}{10pt} GW150914 & $\times$ &  $2815$ & $615$ & $40$ \\
 \hline
 \rule{0pt}{10pt} GW170608 & $\times$ & $1290$ & $303$ & $35$\\
 \hline
\end{tabular}
\caption{Value of the signal-to-noise ratio (SNR) of the considered events for each
detector, taken from \cite{Cardoso:2019rou}.}
\label{table:SNR-events}
\end{table}

\subsection{Detection thresholds in the $(\rho_0,\rho_a)$ plane}
\label{sec:detection-thresholds}

\begin{figure*}
     \centering
     \begin{subfigure}{0.49\textwidth}
         \centering
         \includegraphics[width=\textwidth]{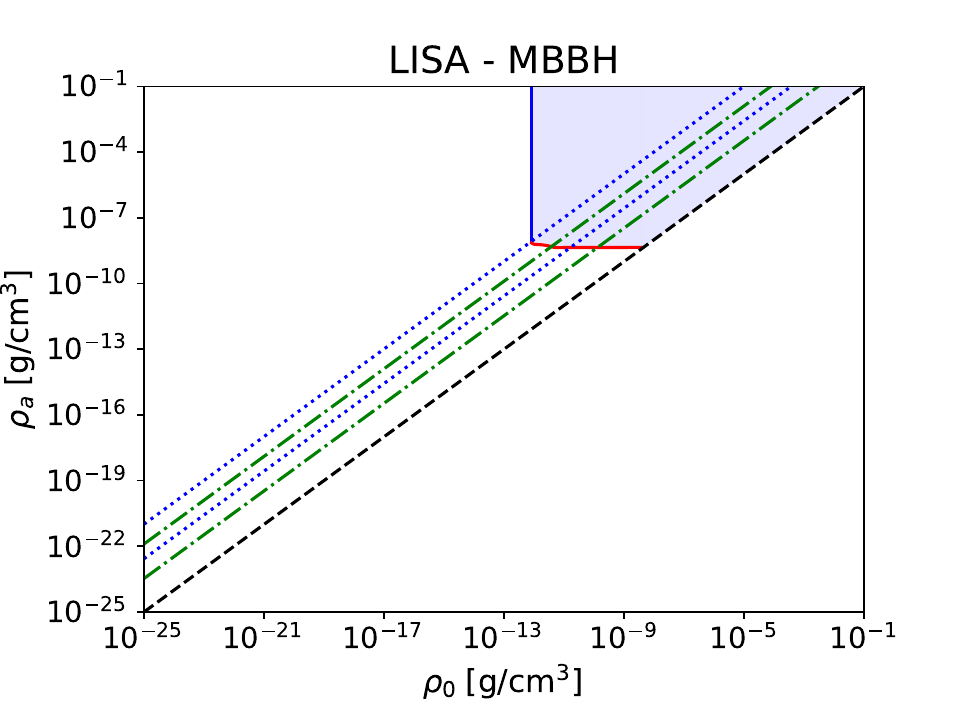}
     \end{subfigure}
     \hfill
     \begin{subfigure}{0.49\textwidth}
         \centering
         \includegraphics[width=0.98\textwidth]{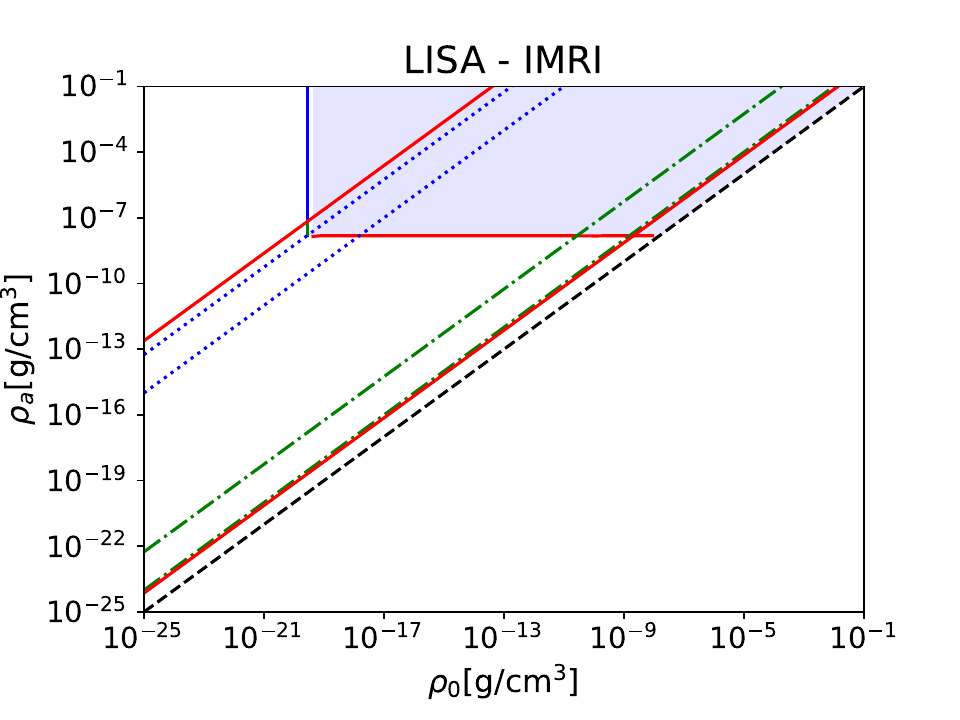}
     \end{subfigure}
     \par\bigskip
     \begin{subfigure}{0.49\textwidth}
         \centering
         \includegraphics[width=\textwidth]{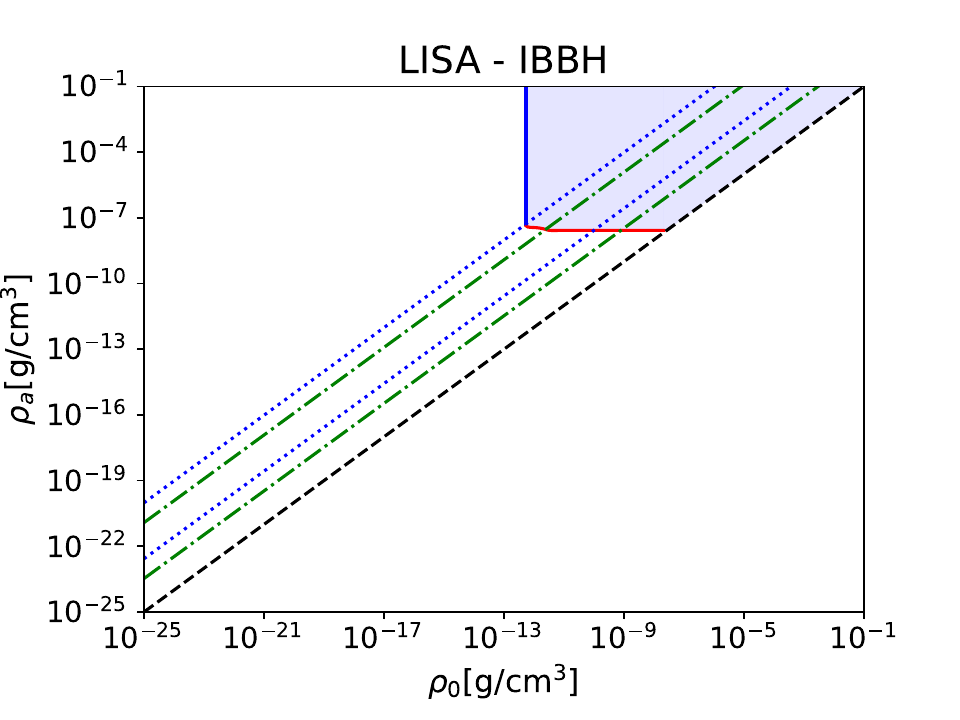}
     \end{subfigure}
     \hfill
     \begin{subfigure}{0.49\textwidth}
         \centering
         \includegraphics[width=1\textwidth]{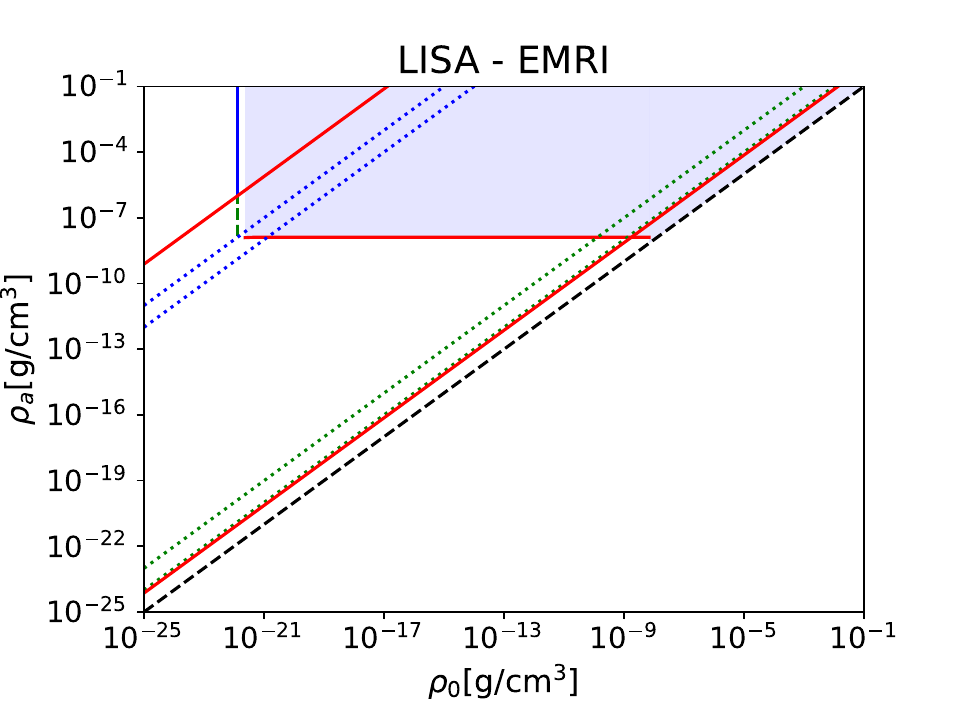}
     \end{subfigure}
	\caption{
	Maps of the detection prospects with LISA for different events, in terms of the dark matter
	parameters	$\rho_0$ and $\rho_a$. The lower right area below the black dashed line is not
	physical. The shaded upper right area shows the region of the parameter space where the dark
	matter environment can be detected.
	}
\label{fig:LISA_results}
\end{figure*}

\begin{figure*}
     \centering
     \begin{subfigure}{0.33\textwidth}
         \centering
         \includegraphics[width=\textwidth]{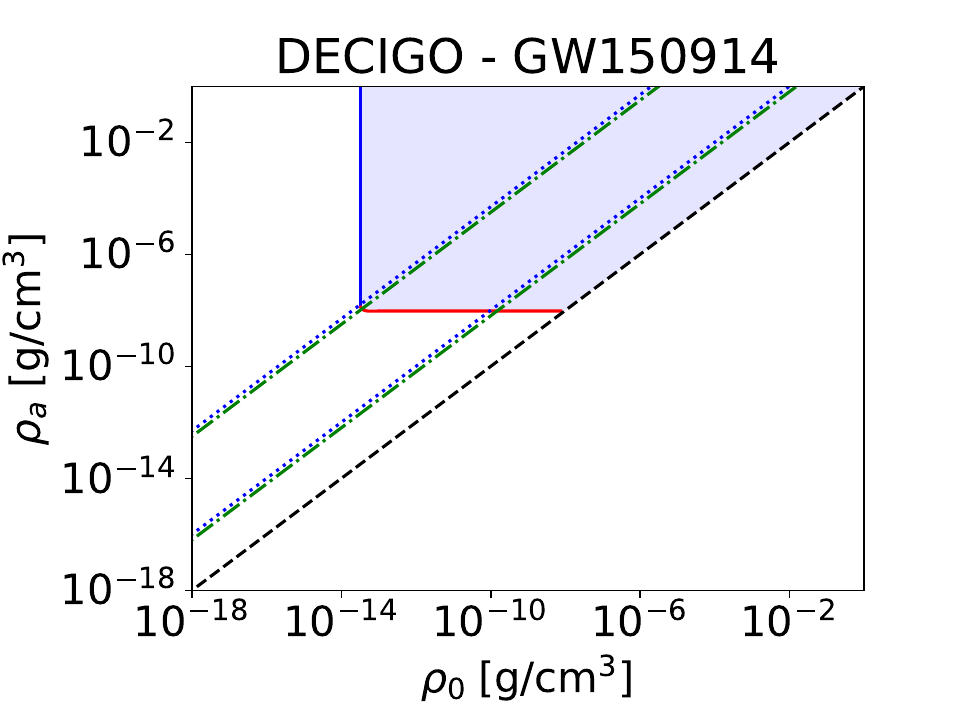}
     \end{subfigure}
     \hfill
     \begin{subfigure}{0.33\textwidth}
         \centering
         \includegraphics[width=0.985\textwidth]{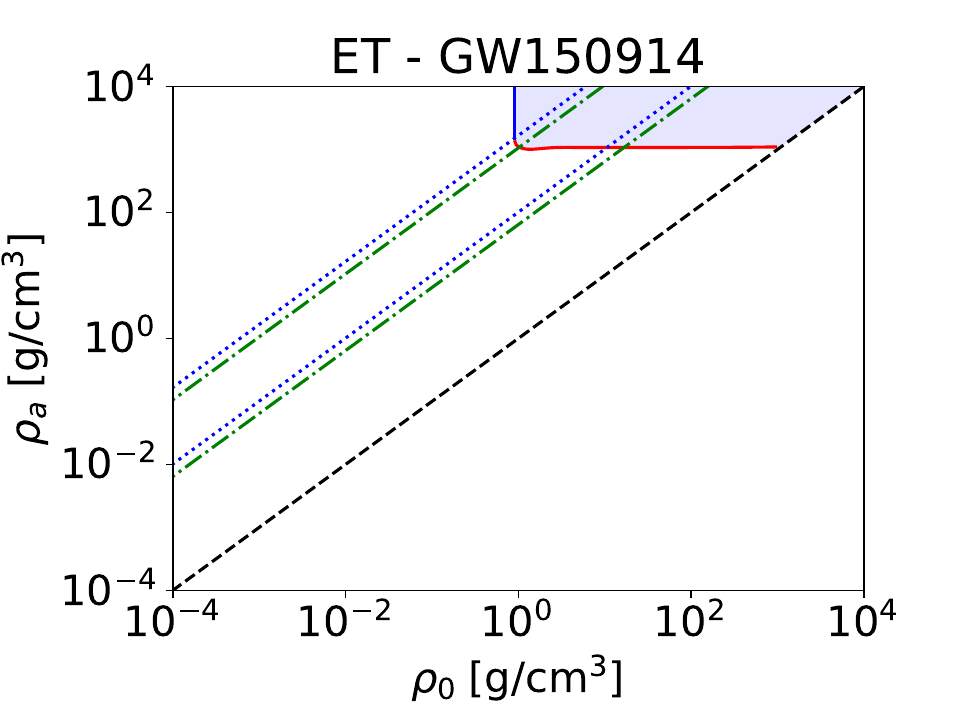}
     \end{subfigure}
     \hfill
     \begin{subfigure}{0.32\textwidth}
         \centering
         \includegraphics[width=0.995\textwidth]{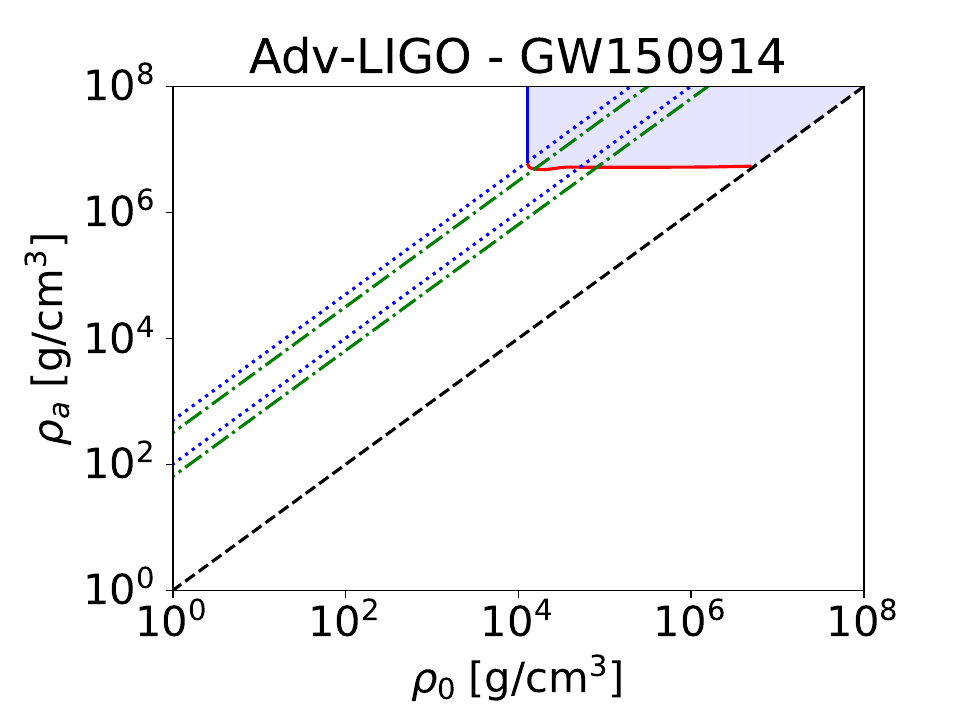}
     \end{subfigure}
     \par\bigskip
     \begin{subfigure}{0.33\textwidth}
         \centering
         \includegraphics[width=\textwidth]{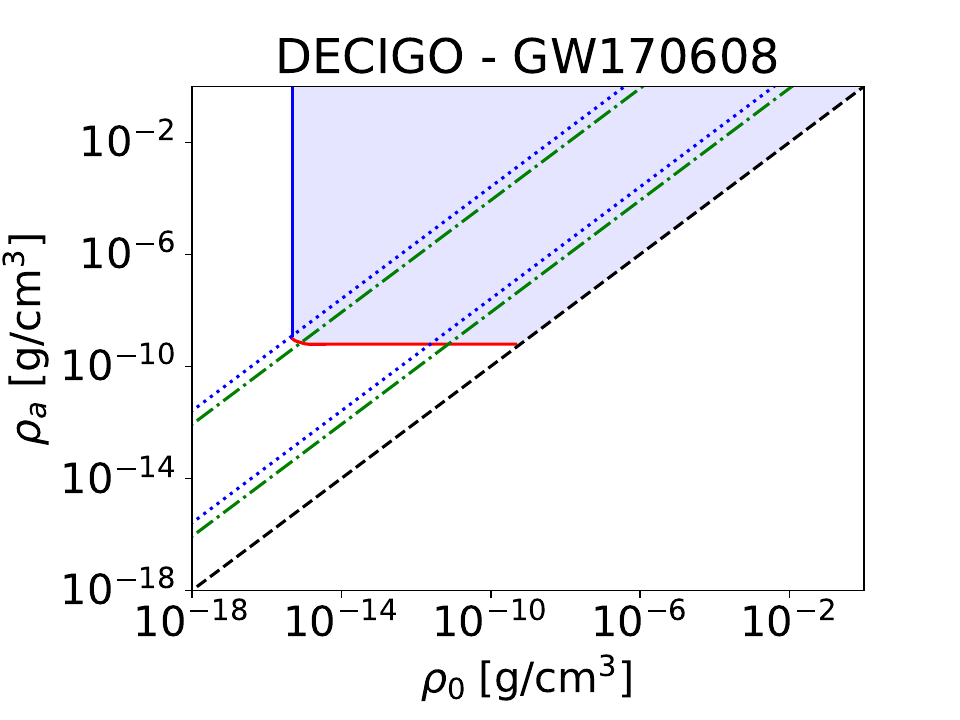}
     \end{subfigure}
     \hfill
     \begin{subfigure}{0.33\textwidth}
         \centering
         \includegraphics[width=0.985\textwidth]{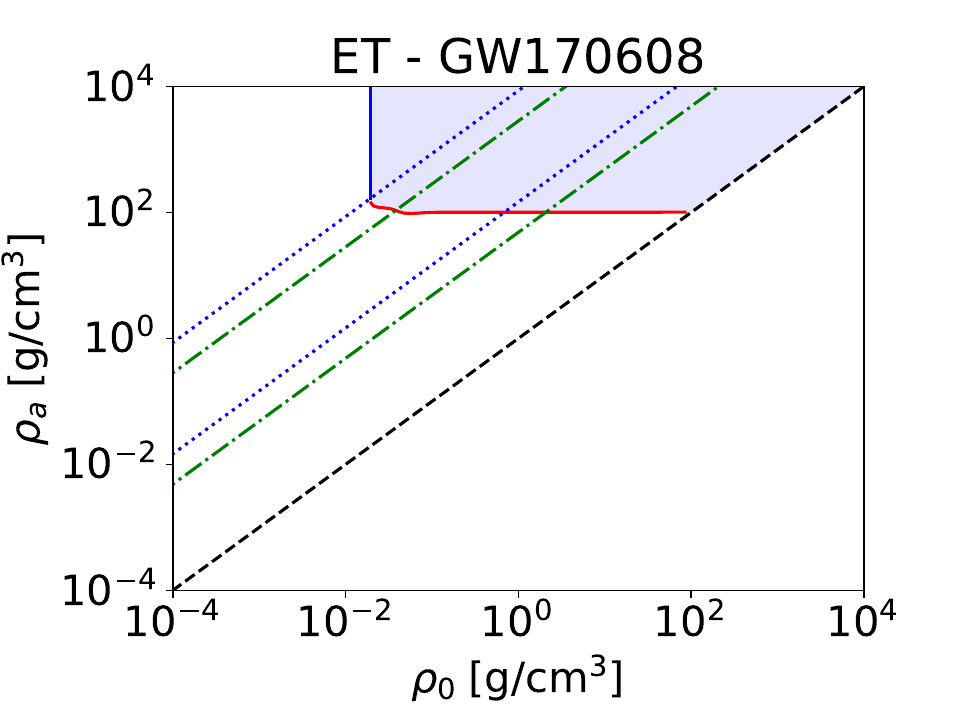}
     \end{subfigure}
     \hfill
     \begin{subfigure}{0.32\textwidth}
         \centering
         \includegraphics[width=0.995\textwidth]{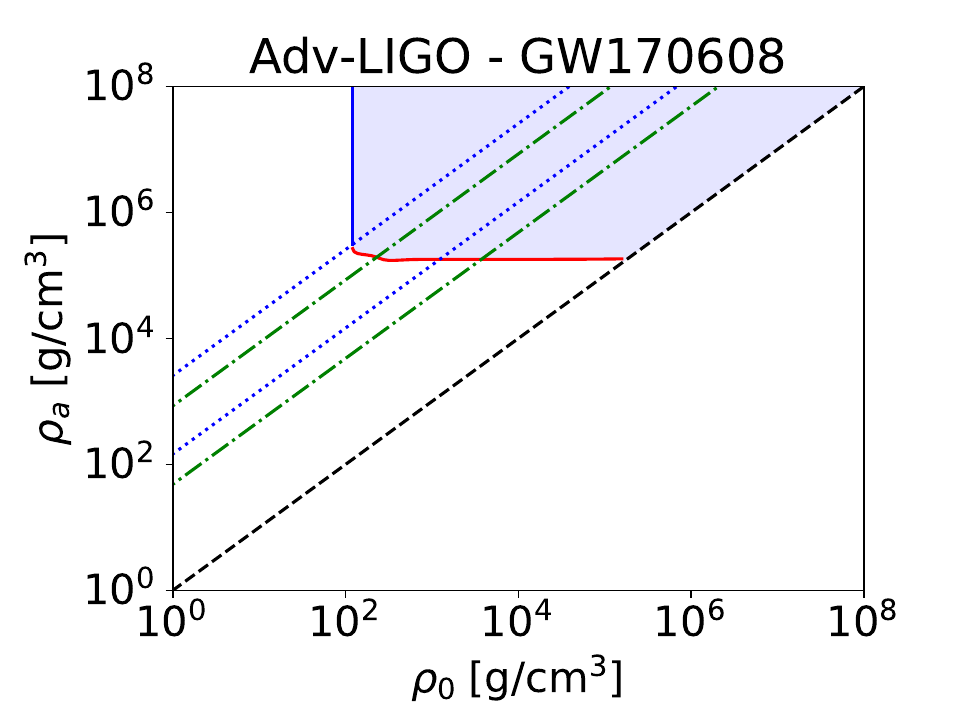}
     \end{subfigure}
    \caption{
    Maps of the detection prospects for three different interferometers
    (from left-to-right: DECIGO, ET, and Adv-LIGO), for the two events GW150914 (upper row)
    and GW170608 (lower row).}
    \label{fig:Detectors_results}
\end{figure*}

We show in Figs.~\ref{fig:LISA_results} and \ref{fig:Detectors_results} our results for the detection
thresholds in the $(\rho_0,\rho_a)$, following the Fisher matrix analysis described in Sec.~\ref{sec:fish}.
Let us first describe the LISA-MBBH case, shown in the upper left panel in Fig.~\ref{fig:LISA_results}.
The lower diagonal black dashed line is the lower limit $y=1$ ($c_s=c$) on the physical part of the parameter
space.
The parallel blue dotted lines are the thresholds $y_{{\rm acc},1}^+$ and $y_{{\rm acc},1}^-$
while the green dot-dashed lines are the thresholds $y_{{\rm acc},2}^+$ and $y_{{\rm acc},2}^-$
(constant-$y$ lines are parallel to the diagonal $y=1$ in the $(\log(\rho_0),\log(\rho_a))$ logarithmic plane).
Because $\nu > 0.16$ there is no dynamical friction.

Then, above the upper blue dotted line $y_{{\rm acc},1}^+$, we are in the large-$y$ regime
(\ref{eq:large-y}) and there is no constraint on $\rho_a$. Thus, we obtain a vertical line
$\rho_0 > \rho_{0\star}$ with $\rho_{0\star} \simeq 8 \times 10^{-13} {\rm g/cm}^3$
This should be compared with the simple estimate (\ref{eq:rho0-star}), which gives
$\rho_{0\star} \gtrsim 10^{-14} {\rm g/cm}^3$ as we have
$f_{\rm min} \simeq 6 \times 10^{-5} \, {\rm Hz}$.
As expected the more accurate Fisher analysis gives a higher value but we roughly recover
the same order of magnitude.
This gives the shaded area to the right of $\rho_{0\star}$ and above the line $y_{{\rm acc},1}^+$
as a region where DM would be detected, mostly because of the accretion contribution $\Psi_{{\rm acc},1}$
on the larger BH.

Between the lines $y_{{\rm acc},1}^+$ and $y=1$, we are in the low-$y$ regime
(\ref{eq:low-y}) where the phase depends on both $\rho_0$ and $\rho_a$.
The Fisher matrix analysis gives an almost flat boundary curve $\rho_a >  \rho_{a\star}$
with $\rho_{a\star} \simeq 5 \times 10^{-9} {\rm g/cm}^3$
This should be compared with the simple estimate (\ref{eq:rhoa-star}), which gives
$\rho_{a\star} \gtrsim 10^{-11} {\rm g/cm}^3$.
Again, the more accurate Fisher analysis gives a higher value but we roughly recover
the same order of magnitude.
In particular, the estimates (\ref{eq:rho0-star}) and (\ref{eq:rhoa-star})
correctly predict the large hierarchy between the thresholds $\rho_{0\star}$ and $\rho_{a\star}$.
This gives the remaining shaded area between the lines $y_{{\rm acc},1}^+$ and $y=1$,
above $\rho_{a\star}$, as a region where DM would be detected, mostly because of the accretion
contribution $\Psi_{{\rm acc},1}$ on the larger BH, but now in the low-velocity self-regulated regime.

The same behaviors are found for the  LISA-IBBH case, shown in the lower left panel
in Fig.~\ref{fig:LISA_results}.
In particular, with $f_{\rm min} \simeq 6 \times 10^{-4} \, {\rm Hz}$,
Eqs.(\ref{eq:rho0-star}) and (\ref{eq:rhoa-star}) give the simple estimates
$\rho_{0\star} \gtrsim 10^{-14} {\rm g/cm}^3$ and $\rho_{a\star} \gtrsim 10^{-9} {\rm g/cm}^3$,
whereas the detailed Fisher matrix inversion gives the more accurate results
$\rho_{0\star} \simeq 5 \times 10^{-13} {\rm g/cm}^3$ and
$\rho_{a\star} \simeq 3 \times 10^{-8} {\rm g/cm}^3$.

Let us now consider the LISA-IMRI case, shown in the upper right panel
in Fig.~\ref{fig:LISA_results}.
In addition to the thresholds $\{y_{{\rm acc},1}^+,y_{{\rm acc},1}^-\}$ and
$\{y_{{\rm acc},2}^+,y_{{\rm acc},2}^-\}$, the red solid lines show the dynamical friction
thresholds $\{y_{{\rm df},2}^+,y_{{\rm df},2}^-\}$.
Above the upper line $y_{{\rm df},2}^+$ we are again in the large-$y$ regime (\ref{eq:large-y}),
with a vertical bound $\rho_{0\star}=3 \times 10^{-20} {\rm g/cm}^3$.
This is again within a factor 100 of the simple estimate  (\ref{eq:rho0-star}), which
gives $\rho_{0\star} \gtrsim 10^{-21} {\rm g/cm}^3$ with
$f_{\rm min} \simeq 6 \times 10^{-3} \, {\rm Hz}$.
In the narrow band $y_{{\rm acc},1}^+ < y < y_{{\rm df},2}^+$ we are in the intermediate regime
(\ref{eq:intermediate-y}), with a weak dependence on $\rho_a$ through $c_s$ in the terms inside
the brackets in Eq.(\ref{eq:Psi-df}). Thus, we still have a roughly vertical line.
Below $y_{{\rm acc},1}^+$ we are in the low-$y$ regime (\ref{eq:low-y}), which is now dominated
by the new dependence of the accretion term on $\rho_a$, which gives a roughly horizontal line
with $\rho_{a\star} \simeq 2 \times 10^{-8} {\rm g/cm}^3$.
The simple estimate (\ref{eq:rhoa-star}) gives $\rho_{a\star} \gtrsim 10^{-9} {\rm g/cm}^3$,
which is again within a factor 100 of the more accurate Fisher matrix result and reproduces
the large hierarchy between $\rho_{0\star}$ and $\rho_{a\star}$.

We obtain similar behaviors for the LISA-EMRI case, shown in the lower right panel
in Fig.~\ref{fig:LISA_results}.
With $f_{\rm min} \sim 3 \times 10^{-3} \, {\rm Hz}$,
the simple estimates (\ref{eq:rho0-star}) and (\ref{eq:rhoa-star})
give $\rho_{0\star} \gtrsim 10^{-24} {\rm g/cm}^3$ and
$\rho_{a\star} \gtrsim 10^{-10} {\rm g/cm}^3$, whereas the more accurate Fisher matrix results are
$\rho_{0\star} \simeq 10^{-22} {\rm g/cm}^3$ and $\rho_{a\star} \simeq 10^{-8} {\rm g/cm}^3$.

We obtain similar behaviors in \ref{fig:Detectors_results} for the DECIGO, ET and Adv-LIGO
detectors, for stellar-mass binaries.
As in the MBBH and IBBH cases, there is no dynamical friction regime.
DECIGO provides constraints on DM environments that are similar to those obtained
from LISA, but the ET and Adv-LIGO cannot detect the dark matter cloud for realistic densities.

Thus, in all cases the detection domain is an upper right region, delimited from the left by
$\rho_{0\star}$, from below by $\rho_{a\star}$, and from the right by the diagonal $\rho_a=\rho_0$.
The simple estimates (\ref{eq:rho0-star}) and (\ref{eq:rhoa-star}) are typically below
the exact thresholds $\rho_{0\star}$ and $\rho_{a\star}$ by a factor of up to 100, but they
reproduce the main trends and the hierarchy between $\rho_{0\star}$ and $\rho_{a\star}$.
The DM detection is dominated by the accretion contribution $\Psi_{\rm acc}$ on the larger BH.
Above the diagonal $y_{{\rm acc},1}^+$, which runs through the lower-left corner of this domain,
the accretion rate is proportional to $\rho_0$ whereas below the diagonal $y_{{\rm acc},1}^+$
it is proportional to $\rho_a$.
Therefore, in the shaded domain above $y_{{\rm acc},1}^+$ we measure $\rho_0$
whereas below $y_{{\rm acc},1}^+$ we measure $\rho_a$.

\setlength{\tabcolsep}{10pt}
\renewcommand{\arraystretch}{1.4}

\begin{table*}
\centering
\begin{tabular}{|l||*{5}{c|}}
 \hline
 \backslashbox{Event}{Detector} & LISA & DECIGO & ET & Adv-LIGO\\ [0.5ex]
 \hline\hline
 \rule{0pt}{5pt} MBBH & $\rho_0 > 8 \times 10^{-13}$ g/cm$^3$ & $\bigtimes$ & $\bigtimes$ &  $\bigtimes$ \\
 \rule{0pt}{5pt}  & $\rho_a > 5 \times 10^{-9}$ g/cm$^3$ & $\bigtimes$ & $\bigtimes$ &  $\bigtimes$ \\
 \hline
 \rule{0pt}{5pt} IBBH & $\rho_0 > 5 \times 10^{-13}$ g/cm$^3$ &$\bigtimes$ & $\bigtimes$ & $\bigtimes$ \\
 \rule{0pt}{5pt}  & $\rho_a > 3 \times 10^{-8}$ g/cm$^3$ & $\bigtimes$ & $\bigtimes$ &  $\bigtimes$ \\
 \hline
 \rule{0pt}{5pt} IMRI & $\rho_0 > 3\times 10^{-20}$ g/cm$^3$ & $\bigtimes$ & $\bigtimes$ & $\bigtimes$  \\
 \rule{0pt}{5pt}  & $\rho_a > 2 \times 10^{-8}$ g/cm$^3$ & $\bigtimes$ & $\bigtimes$ &  $\bigtimes$ \\
 \hline
 \rule{0pt}{5pt} EMRI &   $\rho_0 > 10^{-22}$ g/cm$^3$ & $\bigtimes$ & $\bigtimes$ & $\bigtimes$ \\
 \rule{0pt}{5pt}  & $\rho_a > 10^{-8}$ g/cm$^3$ & $\bigtimes$ & $\bigtimes$ &  $\bigtimes$ \\
 \hline
 \rule{0pt}{5pt} GW150914 & $\bigtimes$ &  $\rho_0 > 3\times 10^{-14}$ g/cm$^3$ &  $\rho_0 > 0.9 $ g/cm$^3$ &  $\rho_0 > 10^4$ g/cm$^3$ \\
 \rule{0pt}{5pt} & $\bigtimes$ &  $\rho_a > 10^{-8}$ g/cm$^3$ &  $\rho_a > 10^3 $ g/cm$^3$ &  $\rho_a > 5 \times 10^6$ g/cm$^3$ \\
 \hline
 \rule{0pt}{5pt} GW170608 & $\bigtimes$ & $\rho_0 > 5 \times 10^{-16}$ g/cm$^3$ &  $\rho_0 > 0.02$ g/cm$^3$ &  $\rho_0 > 120$ g/cm$^3$\\
 \rule{0pt}{5pt} & $\bigtimes$ &  $\rho_a > 10^{-9}$ g/cm$^3$ &  $\rho_a > 101 $ g/cm$^3$ &  $\rho_a > 2 \times 10^5$ g/cm$^3$ \\ \hline
\end{tabular}
\caption{Lower bounds $\rho_{0\star}$ and $\rho_{a\star}$ on the DM density parameters
for a detection of the DM cloud, for various detectors and binary systems.}
\label{tab:rho_a_detect}
\end{table*}


We summarize in Table~\ref{tab:rho_a_detect} the density thresholds $\rho_{0\star}$ and $\rho_{a\star}$
above which the DM cloud can be detected, for the detectors and binary systems
displayed in Figs.~\ref{fig:LISA_results} and \ref{fig:Detectors_results}.
This is only possible at much higher densities than the typical dark matter density on galaxy scales,
which is about $10^{-26}$ to $10^{-23}$ g/cm$^3$
\cite{Navarro:1995iw, Navarro:1996gj, Martinsson:2013ukc, Salucci:2018hqu}.
For comparison, we also note that accretion disks have a baryonic matter density below
$\sim 0.1 \, {\rm g/cm^3}$ for thin disks, and below $10^{-9} {\rm g/cm^3}$ for thick disks
\cite{Barausse:2014tra}, with a lower bound around $10^{-16} {\rm g/cm^3}$.
Therefore, only LISA and DECIGO could detect DM clouds with realistic bulk densities,
$\rho_0 > 10^{-22}$g/cm$^3$ for LISA-EMRI and $\rho_0 > 10^{-15}$g/cm$^3$ for
DECIGO.
The detection of the scalar cloud also requires a very high value of the density parameter $\rho_a$,
 $\rho_a \gtrsim 10^{-8}$g/cm$^3$.
However, this is not the typical density of the DM cloud but only the density close to the Schwarzschild radius,
in the accretion regime regulated by the self-interactions.
On the other hand, DM clouds with densities much higher than typical baryonic accretion disks
may be produced in the early universe, as discussed for instance in
\cite{Berezinsky:2014wya,Brax:2020oye} for several scenarios.
Then, in contrast with the standard CDM case, the dark matter density field would be extremely clumpy,
in the form of a distribution of small and dense clouds (in a manner somewhat similar to primordial
BHs or macroscopic dark matter scenarios, but with larger-size objects).

\subsection{Detection threshold for $\rho_{\rm a}$ and parameter space}
\label{sec:parameter_space}

\begin{figure*}
     \centering
     \begin{subfigure}{0.5\textwidth}
         \centering
         \includegraphics[width=\textwidth]{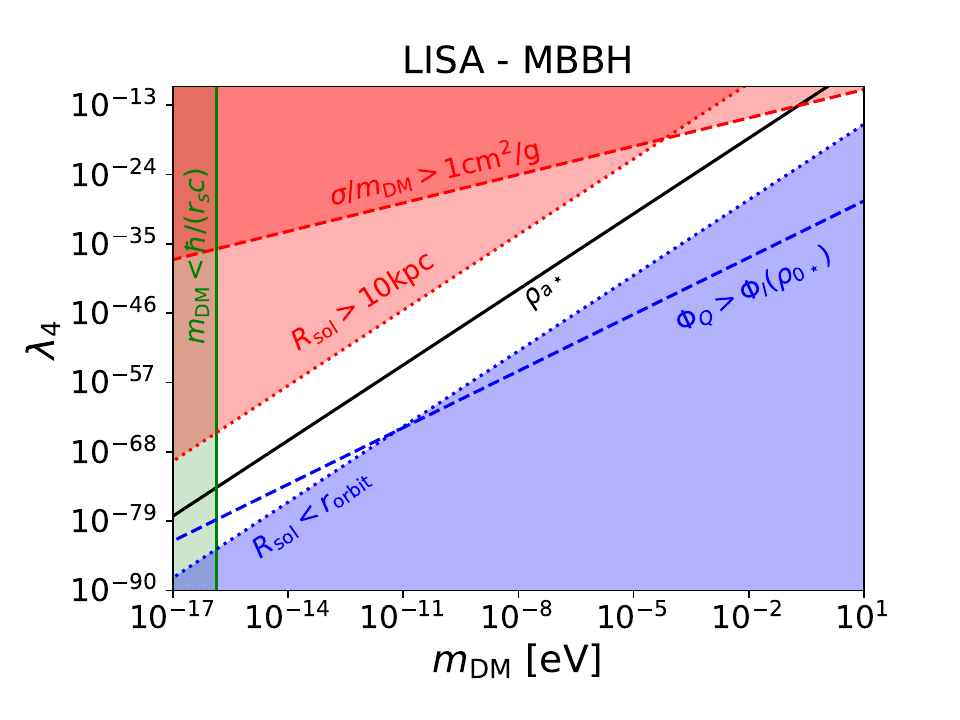}
     \end{subfigure}
     \hfill
     \begin{subfigure}{0.49\textwidth}
         \centering
         \includegraphics[width=\textwidth]{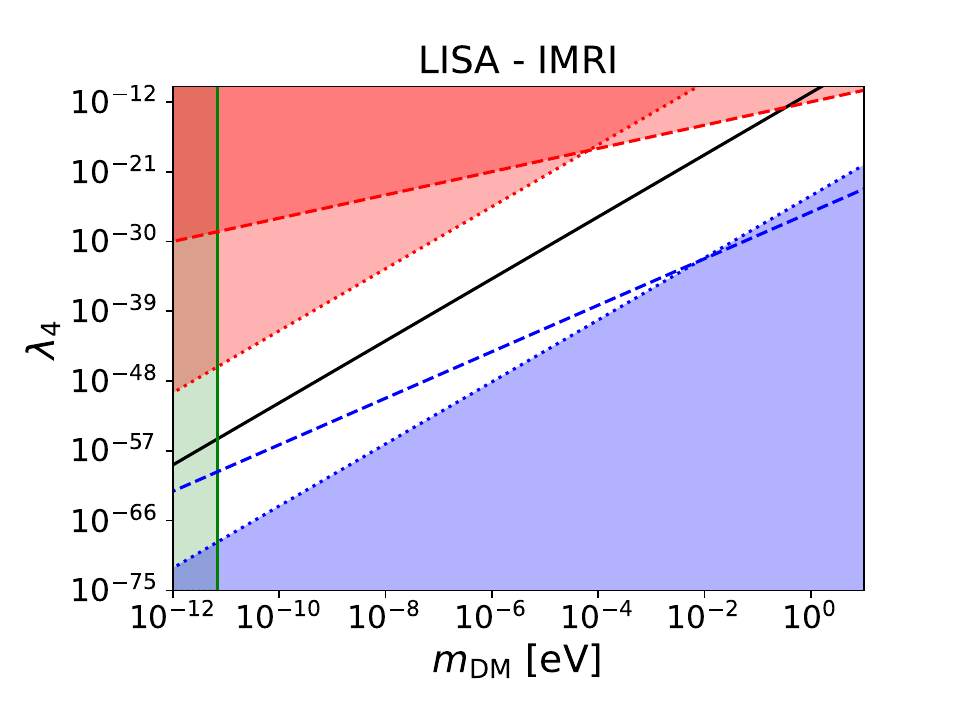}
     \end{subfigure}
     \par\bigskip
     \begin{subfigure}{0.5\textwidth}
         \centering
         \includegraphics[width=1\textwidth]{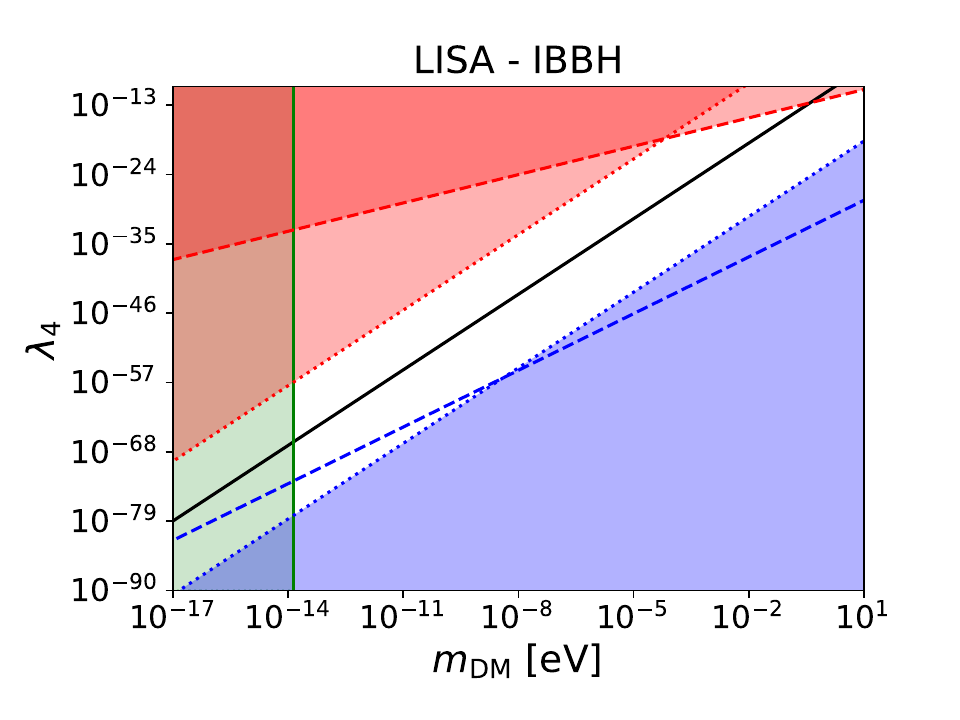}
     \end{subfigure}
     \hfill
     \begin{subfigure}{0.49\textwidth}
         \centering
         \includegraphics[width=\textwidth]{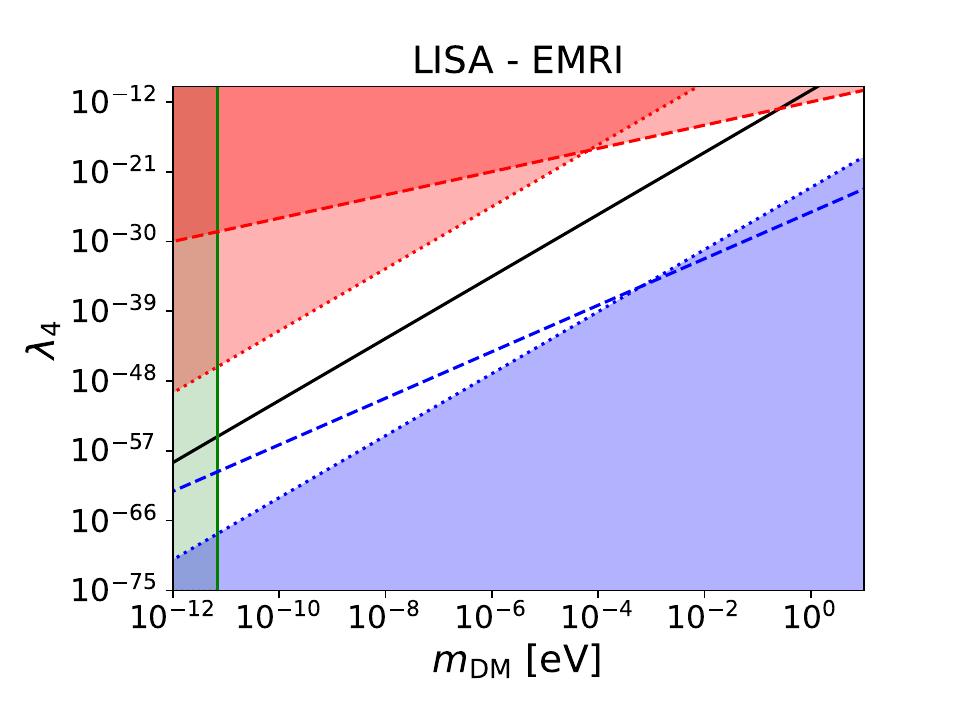}
     \end{subfigure}
    \caption{Domain over the parameter space ($m_{\rm DM}$, $\lambda_4$) where our derivations are applicable, in the case of the LISA interferometer. The white area represents the allowed parameter space.
	The upper left red region is excluded by observational constraints.
	In the lower right blue region the scalar dark matter model is allowed but the assumptions used
	in our computations must be revised.
    The black line corresponds to the detection limit obtained in Fig.~\ref{fig:LISA_results}.
    Parameter values above this line are beyond the detectability range of the interferometer.}
\label{fig:m_lambda_constraints}
\end{figure*}

\begin{figure}
     \centering
     \begin{subfigure}{0.5\textwidth}
         \centering
         \includegraphics[width=\textwidth]{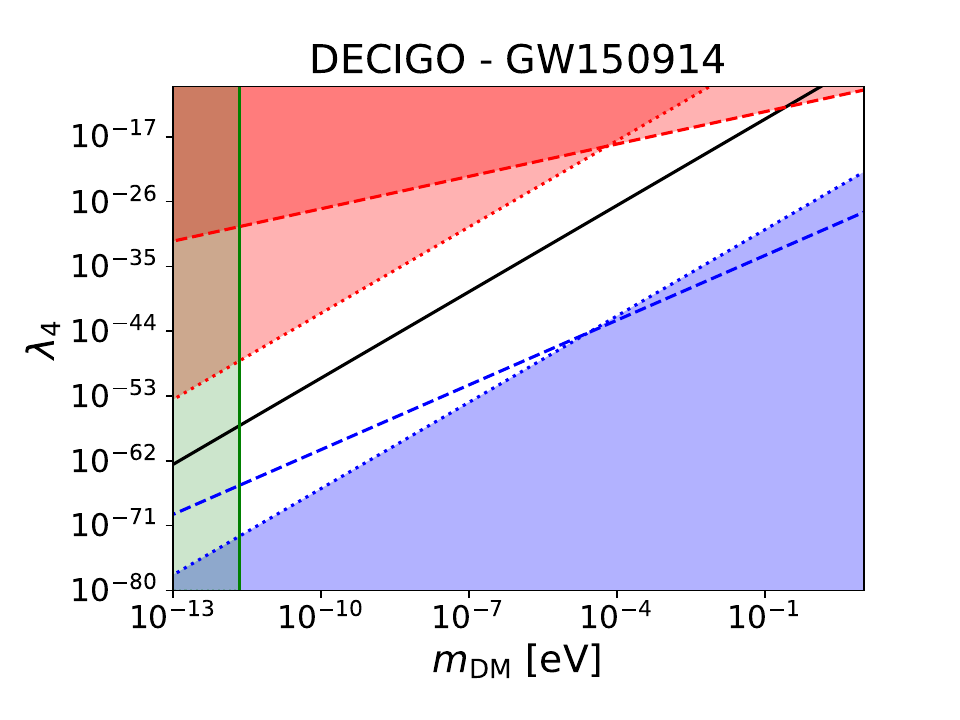}
     \end{subfigure}
     \begin{subfigure}{0.5\textwidth}
         \centering
         \includegraphics[width=\textwidth]{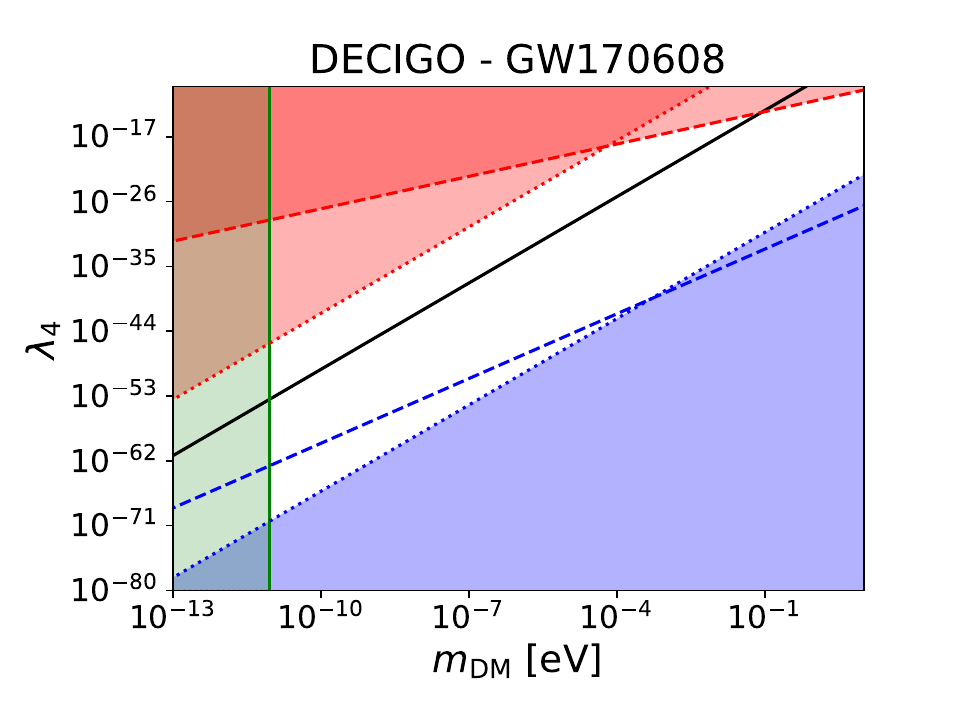}
     \end{subfigure}
          \caption{Domain over the parameter space ($m_{\rm DM}$, $\lambda_4$) where our derivations are applicable
     and detection threshold, as in Fig.~\ref{fig:m_lambda_constraints} but for the interferometer DECIGO.}
     \label{fig:m_lambda_constraints2}
\end{figure}

In this section, we compare the detection threshold $\rho_{a\star}$ obtained in
Table~\ref{tab:rho_a_detect} with the allowed parameter space of our dark matter model,
in the $(m_{\rm DM},\lambda_4)$ plane.
This allows us to check wether this scenario can be efficiently probed by the measurement
of the gravitational waves emitted by BH binary systems embedded in such dark matter
clouds.
Our results are displayed in Figs.~\ref{fig:m_lambda_constraints} and \ref{fig:m_lambda_constraints2}, representing the outcomes for LISA and DECIGO.
We do not consider ET and Adv-LIGO, because they require bulk densities that are probably too
high to be realistic.
Various colored regions on the figures correspond to distinct limits based on either observational constraints or the regime considered in our calculations.

From Eq.(\ref{eq:ra-def}), a detection floor $\rho_{a\star}$ corresponds to an upper ceiling for $\lambda_4$
that scales as $m^4_{\rm DM}$,
\be
\rho_a > \rho_{a\star} : \;\;\; \lambda_4 < \frac{4 m_{\rm DM}^4 c^3}{3 \rho_{a\star} \hbar^3} ,
\label{eq:rho-a-detected-0}
\ee
which reads
\be
\lambda_4 < 3 \times 10^{-19}  \left(\frac{\rho_{a\star}}{1 \, {\rm g/cm^3}}\right)^{-1}
\left(\frac{m_{\rm DM}}{1 \,{\rm eV}}\right)^4 .
\label{eq:rho-a-detected}
\ee
This ceiling is shown by the black solid line labeled $\rho_{a\star}$ that runs through the white area in
Figs.~\ref{fig:m_lambda_constraints} and \ref{fig:m_lambda_constraints2}.

We now describes the constraints that determine the parameter space of the model, with the exclusion
domains shown by the colored regions in the plots.
First, we require the condition (\ref{eq:large-m-rs}), which also reads
\be
m_{\rm DM} > \frac{\hbar c}{2 \NewtonG  m_<} , \;\;\;
m_{\rm DM} > 7 \times 10^{-11} \left( \frac{m_<}{1 \, M_\odot} \right)^{-1} \; {\rm eV} .
\label{eq:no_wavelike_constraint}
\ee
This ensures the validity of the accretion rate (\ref{eq:accretion-rate}) and of the dynamical friction
(\ref{eq:DF}), derived in \cite{Brax:2019npi,Boudon:2022dxi,Boudon:2023qbu} in the large-mass limit
$\partial_r \ll c m_{\rm DM}/\hbar$.
This condition excludes the green area marked by a vertical line on the left in the figures, labeled
$m_{\rm DM}<\hbar/(r_s c)$.

Observations of cluster mergers, such as the bullet cluster, provide an upper bound on the dark matter
cross-section, $\sigma/m_{\rm DM}  \lesssim 1$ cm$^2$/g \cite{Randall:2008}.
This gives the upper bound \cite{Brax:2019fzb}
\be
   \lambda_4 < 10^{-12} \left(\frac{m}{1 \,{\rm eV}}\right)^{\frac{3}{2}} ,
   \label{eq:cross_section_constraint}
\ee
shown by the dashed red line in the upper left corner of the figures, labeled
$\sigma/m_{\rm DM} > 1 \, {\rm cm}^2/{\rm g}$.

Another observational limit, shown by the upper left red dotted line labeled $R_{\rm sol}>10 \, {\rm kpc}$,
is the maximum size of the dark matter solitons. As we wish such solitons to fit inside galaxies,
we require $R_{\rm sol} < 10$ kpc. This gives the upper bound
\be
    \lambda_4 < 0.03 \left( \frac{R_{\rm sol}}{10 \, {\rm kpc}}\right)^2
    \left(\frac{m_{\rm DM}}{1 \, {\rm eV}}\right)^4 .
    \label{eq:small_soliton_constraint}
\ee
This condition is actually parallel to the detection threshold (\ref{eq:rho-a-detected})
and somewhat above it in the Figs.~\ref {fig:LISA_results} and \ref{fig:Detectors_results}.
Therefore, the largest solitons would not be detected by GW.
This will be more clearly seen in Sec.~\ref{sec:radius-constraints} below.

\begin{figure*}
     \centering
     \begin{subfigure}{0.5\textwidth}
         \centering
         \includegraphics[width=0.995\textwidth]{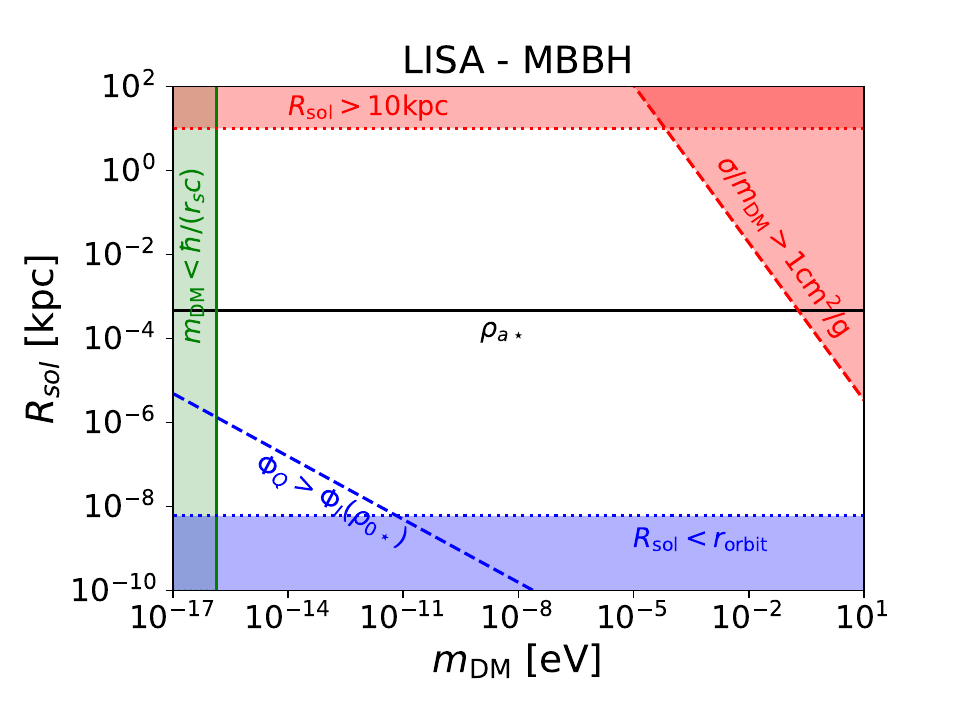}
     \end{subfigure}
     \hfill
     \begin{subfigure}{0.49\textwidth}
         \centering
         \includegraphics[width=1.0\textwidth]{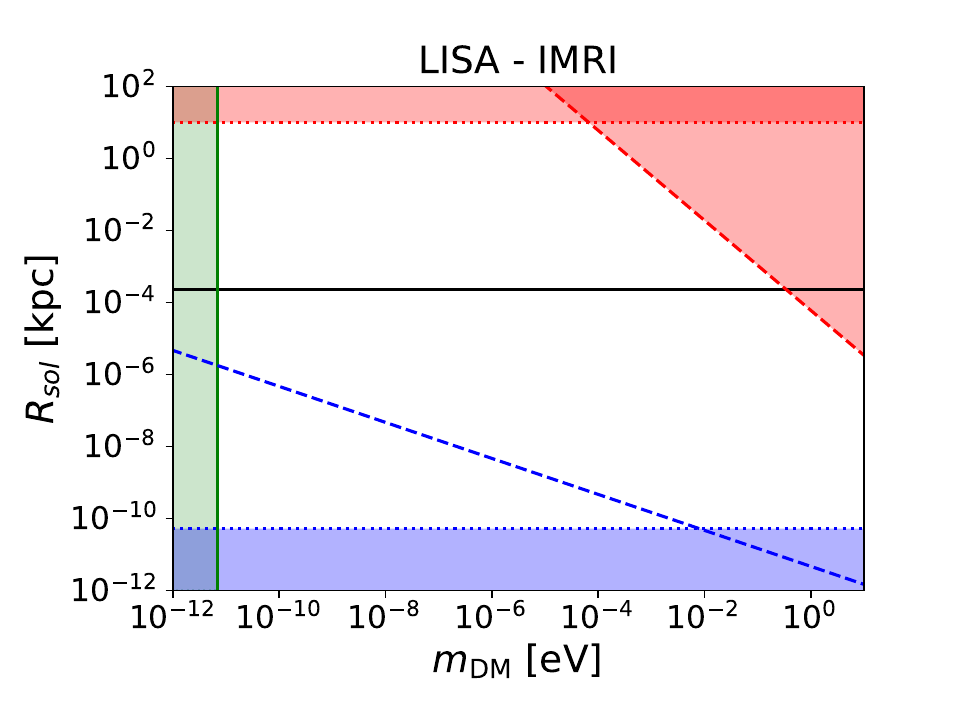}
     \end{subfigure}
     \par\bigskip
     \begin{subfigure}{0.5\textwidth}
         \centering
         \includegraphics[width=1.002\textwidth]{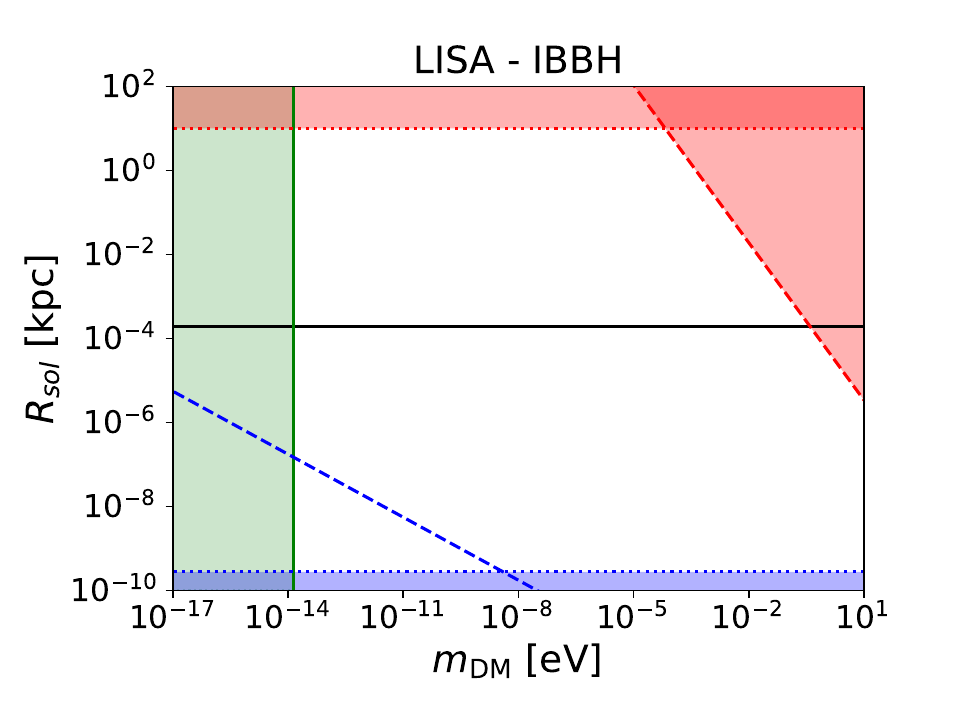}
     \end{subfigure}
     \hfill
     \begin{subfigure}{0.49\textwidth}
         \centering
         \includegraphics[width=1.0\textwidth]{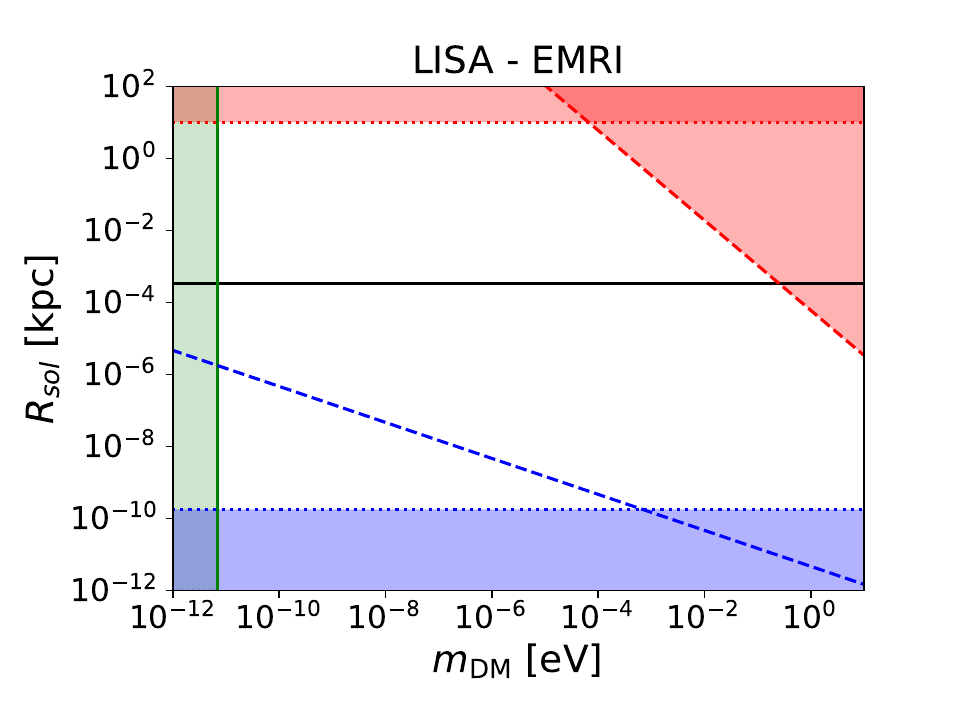}
     \end{subfigure}
    \caption{Domain over the parameter space ($m_{\rm DM}$, $R_{\rm sol})$ where our derivations are
     applicable and detection threshold, in the case of the LISA interferometer
     as in Fig.~\ref{fig:m_lambda_constraints}}
    \label{fig:ra_constraints}
\end{figure*}

\begin{figure}
     \centering
     \begin{subfigure}{0.5\textwidth}
         \centering
         \includegraphics[width=\textwidth]{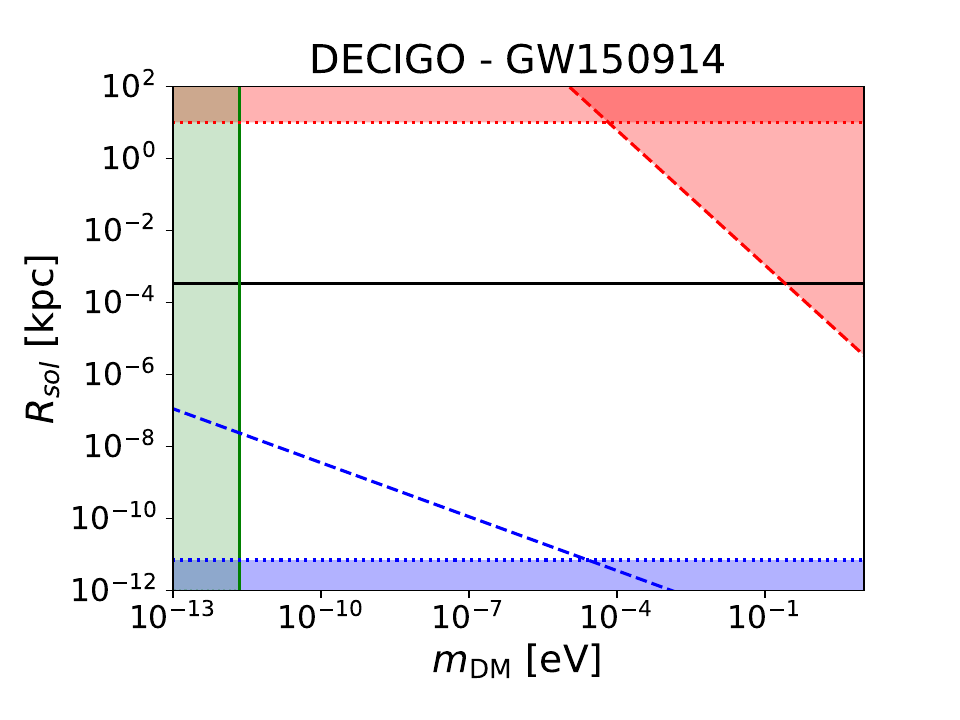}
     \end{subfigure}
     \hfill
     \begin{subfigure}{0.5\textwidth}
         \centering
         \includegraphics[width=1.032\textwidth]{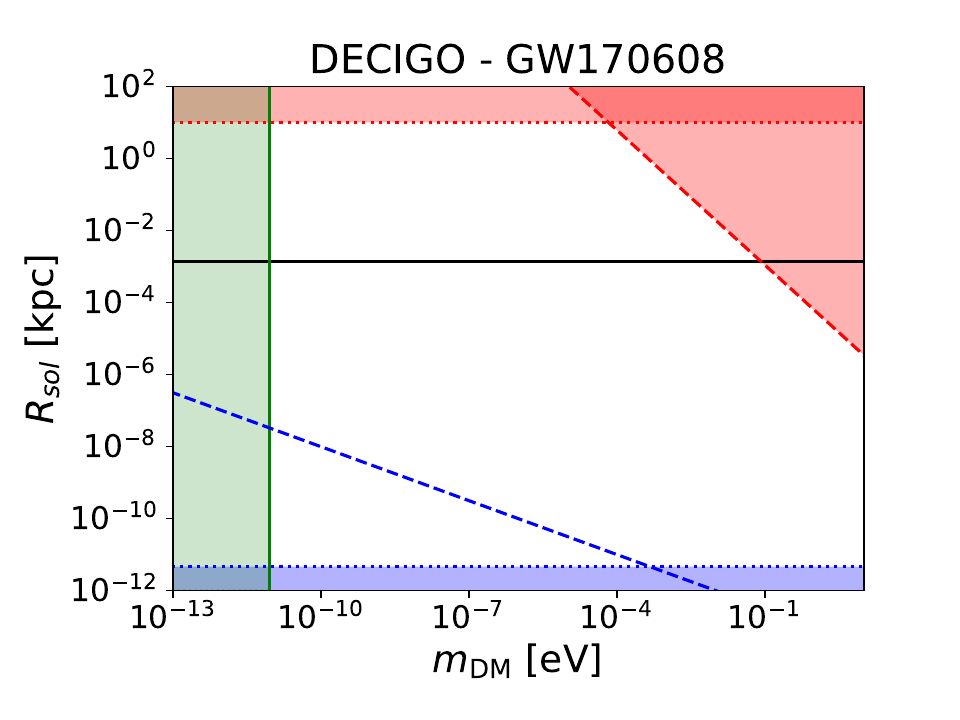}
     \end{subfigure}
     \caption{Domain over the parameter space ($m_{\rm DM}$, $R_{\rm sol})$ where our derivations are
     applicable and detection threshold, in the case of the DECIGO interferometer
	 as in Fig.~\ref{fig:m_lambda_constraints2}.}
	 \label{fig:ra_constraints2}
\end{figure}

Our derivation of the accretion rate (\ref{eq:accretion-rate}) and of the dynamical friction
(\ref{eq:DF}) assumes that the self-interaction dominates over the quantum pressure
\cite{Brax:2019npi,Boudon:2022dxi,Boudon:2023qbu}, in contrast with FDM scenarios where the latter
dominates and the self-interactions are neglected.
The self-interaction potential reads $\Phi_I = c^2 \rho/\rho_a$, whereas the quantum pressure
reads $\Phi_Q = - \hbar^2 \nabla^2\sqrt{\rho}/(2 m_{\rm DM}^2 \sqrt{\rho})$.
This gives the condition $c^2 \rho/\rho_a > \hbar^2 /(r^2 m_{\rm DM}^2)$, where $\rho$ and
$r$ are the density and length scale of interest.
This condition near the BH horizon, with $\rho \sim \rho_a$ and $r \sim r_s$, coincides with the
condition (\ref{eq:no_wavelike_constraint}) and is thus already enforced.
Requiring that this also holds over the bulk of the soliton, at density $\rho_0$ and radius
$r \sim R_{\rm sol}$, gives the additional constraint
\be
\lambda_4 > \frac{8 m_{\rm DM}^3 \sqrt\NewtonG }{3\sqrt{\pi} \hbar^2 \sqrt{\rho_0}} ,
\ee
which reads
\be
\lambda_4 > 6 \times 10^{-38}  \left(\frac{\rho_0}{1 \, {\rm g/cm^3}}\right)^{-\frac{1}{2}}
\left(\frac{m_{\rm DM}}{1 \,{\rm eV}}\right)^3 .
\label{eq:self_bigger_quantum_constraint}
\ee
For the density $\rho_{0\star}$ this is shown by the blue dashed line labeled
$\Phi_Q > \Phi_I(\rho_{0\star})$.
Below this threshold the model itself is not excluded, but our computations should be
be revised as the bulk of the soliton is now governed by the quantum pressure instead of the
self-interactions. However, if the bulk density is greater than $\rho_{0\star}$ this region moves
down to smaller values of $\lambda_4$. Therefore, the blue dashed line is not a strict limit.

Lastly, the area below the blue dotted line labeled $R_{\rm sol} < r_{\rm orbit}$
represents the parameter space where the soliton size
is smaller than the initial orbital radius of the binary system during the measurement.
To ensure the applicability of our calculation across all frequencies, we must thus consider
\be
\lambda_4 > \frac{16 \NewtonG  c m_{\rm DM}^4 r_{\rm orbit}^2}{3 \pi \hbar^3} ,
\ee
which reads
\be
\lambda_4 > 3 \times 10^{-10} \left(\frac{m_{\rm DM}}{1 \, {\rm eV}}\right)^4
\left( \frac{r_{\rm orbit}}{1 \,{\rm pc}}\right)^2 .
\label{eq:rorbit_constraint}
\ee
For $r_{\rm orbit}$ we take the maximum orbital radius, computed with Kepler's third law at
the earliest measurement time, associated with the frequency $f_{\rm obs}(4 \, {\rm yr})$.
This constraint is parallel to the soliton-size condition (\ref {eq:small_soliton_constraint}) and
to the detection threshold $\rho_{a\star}$ in Eq.(\ref{eq:rho-a-detected}).

Hence, the white area in the parameter space indicates where the dark matter model is realistic
and all our calculations apply successfully. More precisely, the upper bounds, associated with the
red exclusion regions, correspond to unphysical regions of the parameter space,
whereas the lower bounds, associated with blue exclusion regions, only correspond to regions
where some of our computations should be revised.
However, where they fall within the detection domain, below the black solid line, it should remain
possible to detect the dark matter environment.

We can see in Fig.~\ref{fig:m_lambda_constraints} and Fig.\ref{fig:m_lambda_constraints2}
that in all cases the detection threshold $\rho_{a\star}$ runs through the white area.
In particular, it is parallel but below the upper bound associated with the soliton size limit and
above the lower bound associated with the orbital radius limit.
Thus, whereas the largest solitons cannot be detected, a large part of the available parameter
space could lead to detection by interferometers such as LISA and DECIGO.
Whereas LISA probes models with a scalar mass $10^{-15} \lesssim m_{\rm DM} \lesssim 1$ eV,
DECIGO is restricted to $10^{-12} \lesssim m_{\rm DM} \lesssim 1$ eV.

\subsection{Constraints on the soliton radius}
\label{sec:radius-constraints}

The two parameters $m_{\rm DM}$ and $\lambda_4$ also determine the soliton size $R_{\rm sol}$,
as seen in Eqs.(\ref{eq:ra-def}) and (\ref{eq:rho-sol-TF}).
As $R_{\rm sol}$ is more relevant for observational purposes than the coupling $\lambda_4$,
we show in Figs.~\ref{fig:ra_constraints} and \ref{fig:ra_constraints2} the application domain
of our computations and the detection threshold $\rho_{a\star}$ in the parameter space
$(m_{\rm DM},R_{\rm sol})$, instead of the plane $(m_{\rm DM},\lambda_4)$ shown
in Figs.~\ref{fig:m_lambda_constraints} and \ref{fig:m_lambda_constraints2} above.

We can see that no experiment can probe galactic-size soltons, $R_{\rm sol} \gtrsim 1$ kpc,
that could be invoked to alleviate the small-scale problems encountered by the standard CDM scenario.
At best, LISA and DECIGO can probe models associated with
$10^{-7} {\rm pc} \lesssim R_{\rm sol} \lesssim 0.1 \, {\rm pc}$. These astrophysical scales range from a percent
of astronomical unit to a tenth of the typical distance between stars in the Milky Way.
Nevertheless, this is still a large fraction of the parameter space.

Scalar dark matter scenarios associated with solitons of such subgalactic size cannot be
constrained by cosmological probes, such as the Lyman-$\alpha$ forest,
or galaxy rotation curves. Their moderate density also evades microlensing detections.
Therefore, their impact on the gravitational waveforms emitted by binary systems
that they could contain would be a key probe of these dark matter scenarios.

\subsection{Comparison with other results}

Our results for the minimal value $\rho_{0\star}$ of the bulk density $\rho_0$ that can be measured
(i.e., its detection threshold) are close to the results obtained in Fig.~2 of \cite{Cardoso:2019rou} from
collisionless dynamical friction, for the DECIGO, ET and ADv-LIGO events and for the LISA interferometer
in the MBBH and IBBH cases,
and somewhat more optimistic than the Bayesian analysis of \cite{santoro2023constraints}.
While, as noticed above, the scalings of the expression (\ref{eq:DF}) for the dynamical friction drag force are quite
general and apply to most media, from collisionless particles to gaseous media and scalar-field dark matter
scenarios, up to some numerical factors, it is not the reason for the similarity in our outcomes. Our main
determinant for the detection threshold, as outlined in Eq.(\ref{eq:rho0-star}), is the accretion drag force, not the
dynamical friction. However, in the high-frequency regime the accretion contribution (\ref{eq:Psi-acc})
shows the same scaling as the dynamical friction contribution (\ref{eq:Psi-df}),
$\Psi \sim (\NewtonG^3 {\cal M}^2 \rho_0/c^6) (\pi \NewtonG {\cal M} f/c^3)^{-16/3}$, up to numerical factors
and ratios of the binary masses. This explains why we recover similar results to those of
Fig.~2 of \cite{Cardoso:2019rou} for the cases where the binary masses are similar and those mass ratios
are of the order of unity.

However, for the IMRI and EMRI cases with the LISA interferometer, our findings are more promising
as we obtain detection thresholds that are lower by factors  $\sim 10^3$ as compared with
Fig.~2 of \cite{Cardoso:2019rou}. This is because the accretion contribution (\ref{eq:Psi-acc})
is greater than the dynamical friction contribution (\ref{eq:Psi-df}) that would be associated with the most massive BH by a factor
$m_i^2/(m\mu) \sim m_>/m_< \sim 1/\nu$, which reaches $10^3$ and $10^4$ for IMRI and EMRI.

Here we note that our results (\ref{eq:Psi-acc}) and (\ref{eq:Psi-df}) actually differ
from the Bondi-accretion and collisionless dynamical friction results of \cite{Cardoso:2019rou}
by powers or $\nu$, which are relevant in case of IMRI and EMRI.
As compared with \cite{Cardoso:2019rou}, our contribution from the accretion drags is enhanced
by the factor $2m_i^2/(m\mu) \sim 2 m_>/m_< \sim 2/\nu$ associated with the accretion onto
the more massive BH.
This term originates from the factor $\dot m/m$ in Eq.(\ref{eq:dot-e-dot-a}), which
comes from the drift of the Runge-Lenz vector (\ref{eq:dAdt-general}). It seems that the
expressions used in \cite{Cardoso:2019rou} only take into account the term $2\dot\mu/\mu$
in the accretion drag, that is, the accretion contribution to the force $F(t)$ in
Eqs.(\ref{eq:ddot-r-F}) and (\ref{eq:dot-e-dot-a}), and neglect the factor $\dot m/m$.

Our contribution (\ref{eq:Psi-df}) for dynamical friction shows the same scalings as in
\cite{Cardoso:2019rou}. However, as we only include the contribution from the smaller BH,
because of the frequency thresholds, its value is reduced because of the terms $m_i^3/(\mu^2 m)$,
which yield a suppression factor $\sim \nu^3$ for small $\nu$.
This is because we consider a self-interacting scalar-field environment instead of
collisionless particles.
This shows the possible impact of the properties of the medium on the gravitational drag.
However, in our case this term is subdominant as compared with the accretion contribution
and it may be difficult to estimate its precise value from observations.

Our detection thresholds are much lower than those shown in Fig.~1 of \cite{Cardoso:2019rou} for
collisionless accretion. This is because the accretion of scalar field dark matter if much more efficient
than that of collisionless particles (but less efficient than that of a perfect gas at low Mach numbers),
see \cite{Boudon:2022dxi,Boudon:2023qbu}. Indeed, pressure forces restrict tangential motion
and funnel particles in the radial direction \citep{Shapiro:1983du}.
This also gives a different velocity and frequency dependence for the accretion drag associated with
collisionless and self-interacting dark matter.

\section{Conclusion}
\label{sec:conclusion}

The detection of GWs has already given important results for fundamental physics, e.g. the near equality between the speed of GWs and the speed of light \cite{LIGOScientific:2017bnn, LIGOScientific:2019fpa, Liu:2020slm}. In this paper, we suggest that future experiments could reveal some key properties of dark matter. As an example, we focus on scalar dark matter with quartic self-interactions and assume that the dark matter density of the Universe is due to the misalignment mechanism for the scalar field. Locally inside galaxies, these models can give rise to dark matter solitons of finite size where gravity and the repulsive self-interaction pressure balance exactly. This regime applies when the size of the solitons is much larger than the de Broglie wavelength of the scalar particles. In this case, these solitons could be pervasive in each galaxy and BHs could naturally be embedded within these scalar clouds when inspiralling towards each other in binary systems.
The scalar clouds have three effects on the orbits of the binary systems.
First, the gravity of the cloud modifies the trajectories of the BHs.
Second, dark matter accretes onto the BHs and slows them down.
Third, in the supersonic regime the dynamical friction due to the gravitational interaction between the
BHs and distant streamlines further slows them down.
These effects can lead to significant deviations of the binary orbits and therefore to perturbations of the GW signal emitted by the pair of BHs.
The cloud gravity gives a -3PN contribution to the gravitational waveform.
The accretion gives a -4PN or -5.5PN effect at low or high frequency,
whereas the dynamical friction gives a -5.5PN contribution.
As such, these effects are not degenerate with the relativistic corrections that
appear at higher post-Newtonian orders.

For a large part of the scalar dark matter parameter space, future experiments such as LISA
and DECIGO should be able to observe the impact on GW of these dark matter environments,
provided binary systems are embedded within such scalar clouds.
This would give new clues about the nature of dark matter.
Within the framework of the scalar field models with quartic self-interactions studied in this paper,
this would give indications on the value of the bulk dark matter density $\rho_0$ as well as the characteristic
density $\rho_a$ of Eq.(\ref{eq:ra-def}), that is, the combination $m_{\rm DM}^4/\lambda_4$.
This would also give an indirect estimate of the size $R_{\rm sol}$ of the solitons,
from Eq.(\ref {eq:rho-sol-TF}).
The relatively high values of $\rho_0$ required for detection, at least a few hundred times
above the dark matter density in the Solar system for EMRI with LISA, suggest that this probe
is mostly relevant for scenarios where the scalar clouds form at high redshifts,
giving rise to a very clumpy dark matter distribution.
The fact that we have not detected such dark matter effects in the ET and LIGO events
is consistent with the high bulk densities, $\rho_0 \gtrsim 1 \, {\rm g/cm}^3$, that are needed
to allow a detection with these interferometers.

On the other hand, the scenarios that can be probed through their impact on binary GW waveforms,
studied in this paper, correspond to small clouds below $0.1$ pc that cannot be constrained
by cosmological probes or galaxy rotation curves, while there density is still too small
to be detected by microlensing. Therefore, GW waveforms would be a key probe of these
dark matter models.

Perturbations to the gravitational waveforms may result from diverse environments, including
gaseous clouds or dark matter halos associated with other dark matter models.
In all cases where such environments are present, we can expect accretion and dynamical friction
to occur and slow down the orbital motion.
It would be interesting to study whether one can discriminate between these different environments.
As shown in this paper, to do so we could use the magnitude of these two effects and also
the parts in the data sequence where dynamical friction appears to be active or not.
Indeed, depending on the medium dynamical friction is expected to be negligible in some
regimes, such as subsonic velocities. If one can extract such conditions from the data,
one may gain some useful information on the environment of the binary systems.
We leave such studies to future works.

\section*{Aknowledgments}
A.B. would like to thank Andrea Maselli for his help in the first stage of this project.

\bibliography{GW}

\end{document}